\documentclass[11pt]{article}%
\usepackage{hyperref}
\usepackage{latexsym}
\usepackage{amssymb,amsfonts,amsmath}
\usepackage{graphicx}
\usepackage{indentfirst}
\usepackage{amsmath}
\usepackage{amsfonts}
\usepackage{amssymb}%
\pdfoutput=1
\makeatletter

\@addtoreset{equation}{section}
\makeatother
\ifx\pdfoutput\undefined
\else
\fi
\hypersetup{colorlinks=false,bookmarksopen,bookmarksnumbered,citecolor=blue,
pdfstartview=FitH}

\parskip=\medskipamount

\def\a{\alpha}

\def\b{\beta}

\def\g{\gamma}

\newcommand{\be}{\begin{equation}}
\newcommand{\ee}{\end{equation}}
\newcommand{\bea}{\begin{eqnarray}}
\newcommand{\eea}{\end{eqnarray}}

\newcommand{\ba}{\begin{array}}
\newcommand{\ea}{\end{array}}

\def\double #1{#1{\hbox{\kern-2pt $#1$}}}

\newcommand{\bsubeq}{\begin{subequations}}
\newcommand{\esubeq}{\end{subequations}}

\begin{document}

\begin{titlepage}
\begin{flushright}
DISIT-2017\\
YITP-17-68\\
ARC-17-09
\par\end{flushright}
\vskip 1.5cm
\begin{center}
\textbf{\Large \bf Wess-Zumino and Super Yang-Mills Theories  \\ 
in  D=4 Integral Superspace}
\vskip 1.5cm
{\large
L. Castellani$^{~a,b,c,}$\footnote{leonardo.castellani@uniupo.it},
R. Catenacci$^{~a,c,d,}$\footnote{roberto.catenacci@uniupo.it},
and
P.~A. Grassi$^{~a,b,c,e,}$\footnote{pietro.grassi@uniupo.it}
\medskip
}
\vskip 0.5cm
{
\small\it
\centerline{${(a)}$ Dipartimento di Scienze e Innovazione Tecnologica, Universit\`a del Piemonte Orientale}
\centerline{\it Viale T. Michel, 11, 15121 Alessandria, Italy}
\medskip
\centerline{${(b)}$ {\it INFN, Sezione di Torino, via P. Giuria 1, 10125 Torino, Italy} }
\medskip
\centerline{${(c)}$ {\it Arnold-Regge Center, via P. Giuria 1, 10125 Torino, Italy}}
\medskip
\centerline{${(d)}$ {\it Gruppo Nazionale di Fisica Matematica, INdAM, P.le Aldo Moro 5, 00185 Roma, Italy} }
\medskip
\centerline{${(e)}$ \it Center for Gravitational Physics, Yukawa Institute for Theoretical Physics,}
\centerline{\it Kyoto University, Kyoto 606-8502, Japan}
}
\par\end{center}
\vfill{}
\begin{abstract}
{We reconstruct the action of $N=1, D=4$ Wess-Zumino and 
$N=1, 2, D=4$ super-Yang-Mills theories, using integral top forms on the supermanifold 
${\cal M}^{(4|4)}$. Choosing different Picture Changing Operators, we show the 
equivalence of their rheonomic and superspace actions.  The corresponding supergeometry 
and integration theory are discussed in detail. This formalism is an efficient 
tool for building supersymmetric models in a geometrical framework. 
}
\end{abstract}
\vfill{}
\vspace{1.5cm}
\end{titlepage}
\newpage\setcounter{footnote}{0}
\tableofcontents
\section{Introduction}

In some recent papers
\cite{Castellani:2014goa,Castellani:2015paa,Castellani:2015ata}, we explored
the role of the supermanifolds and their integration theory for
applications to gauge theories, supergravity and string theories. 

The superspace technique has been 
invented to describe supersymmetric theories with manifestly supersymmetric actions. This is achieved 
by adding fermionic coordinates to the bosonic manifold and using 
Berezin integration. Nonetheless the geometric point of view needs further clarification.
During the recent years, due to progress in fundamental string theory \cite{berkovits,Witten:2012bg} and 
due to progress in the understanding of integration theory on supermanifolds (see for ex. \cite{voronov-book,Witten:2012bg}), a more solid and 
fruitful framework for superspace actions has been built.  

A convenient way to
write a supersymmetric action in superspace, as an integral of integral forms
on supermanifolds $\mathcal{M}^{(n|m)}$, is the following
\begin{equation}
S=\int_{\mathcal{M}^{(n|m)}}\mathcal{L}^{(n|0)}\wedge{\mathbb{Y}}%
^{(0|m)}\label{IntroA}%
\end{equation}
where the Lagrangian $\mathcal{L}^{(n|0)}$ is a superfield $(n|0)$ superform and
${\mathbb{Y}}^{(0|m)}$ is a \textit{Picture Changing Operator}
(PCO), or using a proper mathematical identification (see for ex. \cite{Castellani:2017ycm}), is the Poincar\'e dual form of the 
embedding of a $n$-dimensional bosonic submanifold into the supermanifold ${\cal M}^{(n|m)}$. 
${\mathbb{Y}}^{(0|m)}$  belongs to the super de Rham cohomology $H^{(0|m)}(\mathcal{M}%
^{(n|m)})$. 

The choice of the PCO determines the representation for the
supersymmetric theory: the simplest PCO 
constructed in terms of the fermionic coordinates $\theta^{\alpha}$ (with
$\alpha=1,\dots m$), and their corresponding one-forms $\psi^{\alpha}%
=d\theta^{\alpha}$, is given by $\theta^{m}\delta^{m}(\psi)$. When inserted into (\ref{IntroA}) it 
reproduces the component action. When instead a supersymmetric PCO is used, it yields a superfield action with manifest supersymmetry.
Different choices of the PCO's produce different representations of the same theory with 
different amounts of manifest supersymmetry and passing from one to another leads to equivalent theories when the PCO's differ by exact terms and $\mathcal{L}^{(n|0)}$ is closed. 

The purpose of the present paper is to study the four
dimensional case, with different amounts of supersymmetry. In particular, we will study
the case $N=1$ and $N=2$. The cases $D=1$, $D=2$ and $D=3$ are treated in \cite{Castellani:2017ycm}, \cite{d2}, \cite{3dsuper}.

The paper is organized as follows: 
\begin{enumerate}
\item In Sec. 2, we review the Wess-Zumino model for a chiral field from the superspace point of view. 
This is the usual construction of the textbooks and we use it to set the stage. Then, we consider the 
geometric formulation of the {\it rheonomic} formalism. That framework uses only geometric 
ingredients: superforms, exterior differential and wedge product. Finally, we rewrite the action 
using the integral form formulation which projects the geometric action to the superspace action. 
We perform the computation explicitly to illustrate all steps and we postpone the mathematical 
construction of the PCO in later sections. 
\item In Sec. 3, we review the $N=1$ super Yang-Mills theory in the superspace framework. 
Differently from the usual prepotential construction (see for example the textbook \cite{baggerwess,GGRS}), suitable only for $D=4$, we use the form language (see \cite{Castellani}) and we discuss the solution 
of the constraints. This allows us to write both the superspace action and the 
geometric action in terms of the 
gaugino field strength $W^\a, \bar W^{\dot \a}$. The dependence of the geometric action upon the 
rigid gravitinos $\psi^\a, \bar \psi^{\dot \a}$ admits a straightforward generalization to supergravity couplings 
and it encodes all possible information. The geometric action is built and the equations of motion 
are given. Finally, we explore two possible choices of the PCO's leading either to the component 
action or to the well-kwown superspace action. 
\item As a further example, in Sec. 4 we consider the case of $N=2$ super-Yang-Mills. We briefly review the $N=2$ superspace action (which consists of only one term integrated 
over the full superspace) and we discuss the rheonomic action. In the long and complicated
rheonomic action displayed in the textbook \cite{Castellani}, only one term is relevant in order
to reproduce the superspace action. The relation with the $N=2, D=4$ action is achieved by 
changing the PCO, using the closure of the rheonomic action. 
\item In Sec. 5, we summarize the mathematical aspects of the derivation. We review the structure 
of the integral superspace, considering the full complex of integral forms and of superforms. We 
review the action of different operators and the notion of picture number. An important 
issue is the Lorentz symmetry for integral forms, discussed in Sec. 5.2. The 
volume forms and the PCO's are built in the subsequent sections with detailed derivations. 
The final two theorems are needed for the supergravity extension of the present framework.\end{enumerate}

\section{D=4 N=1 Integral Wess-Zumino Model}

It is important to clarify the integral form formulation of the most
well-known example of supersymmetric model, namely the Wess-Zumino model. It
describes a chiral multiplet and the field content is given by a complex scalar $\phi$,
two fermions $\lambda^{\alpha}, \bar\lambda^{\dot\alpha}$ and a complex
auxiliary field $f$. The auxiliary field $f$ guarantees the closure of the
off-shell supersymmetry. On shell, $f$ is set to zero and the degrees of freedom
of the fermions are halved by the equations of motion, so that they match the
bosonic degrees of freedom.

In Sec 2.1 we review the superspace action in the conventional
Weyl/anti-Weyl notation. We give the action in
component fields. In Sec 2.2 we review the geometric (rheonomic) action
described in the book \cite{Castellani}, rewriting it into chiral notation. In Sec. 2.3 we construct the
action on the supermanifold $\mathcal{M}^{(4|4)}$ and show how to reproduce the
superspace action and the component action. For that we need suitable PCO's to
project the geometric action along different supersymmetry realizations. The relevant PCO's
will be described later in Sec. 5. 

\subsection{WZ superspace action}

The spinors are taken in the Weyl/anti-Weyl representation in order to compare
our formulas with the usual D=4 N=1
superspace \cite{GGRS,baggerwess}. In that framework the supermultiplet is
described by a single complex superfield $\Phi(x,\theta,\bar{\theta})$
satisfying
\begin{equation}
\bar{D}_{\dot{\alpha}}\Phi=0\,. \label{BWA}%
\end{equation}
where $\bar{D}_{\dot{\alpha}} = \partial_{\bar \theta^{\dot \a}} - i \theta^\a \partial_{x^{\a\dot\a}}$ 
(see also Sec.~5.1 for notations, differential operators and their algebra). 
Equations (\ref{BWA}) are easily solved by introducing the chiral coordinates $(y^{\alpha\dot{\alpha}}\equiv
x^{\alpha\dot{\alpha}}-i\theta^{\alpha}\bar{\theta}^{\dot{\alpha}}%
,\theta^{\alpha},\bar{\theta}^{\dot{\alpha}})$. The chiral
superfield $\Phi$ is independent of $\bar{\theta}$ and can be decomposed
as follows
\begin{align}
\Phi(y,\theta)  &  =\phi(y)+\lambda_{\alpha}(y)\theta^{\alpha}+f(y)\frac
{\theta^{2}}{2}\label{BWBA}\\
&  =\phi+\lambda_{\alpha}\theta^{\a}+(\frac{1}{2}f\theta^{2}%
-i\theta^{\alpha}\bar{\theta}^{\dot{\alpha}}\partial_{\alpha\dot{\alpha}}%
\phi)-\frac{i}{2}\theta^{2}\bar{\theta}^{\dot{\alpha}}\partial_{\dot{\alpha
}\alpha}\lambda^{\alpha}+\frac{1}{8}\theta^{2}\bar{\theta}^{2}\partial^{2}%
\phi\,.
\end{align}
where $\theta^2 = \theta^{\alpha}\epsilon
_{\alpha\beta}\theta^{\beta}$, $\bar \theta^2 = \bar{\theta}^{\dot{\alpha}}\epsilon
_{\dot{\alpha}\dot{\beta}}\bar{\theta}^{\dot{\beta}}$, and the components $\phi,\lambda^{\alpha}$ and $f$ in the last line depend
on $x$. The free equations of motion (we comment later on the introduction of
a superpotential) are
\begin{equation}
\bar{D}^{2}D^{2}\Phi=0\,. \label{BWB}%
\end{equation}
In components they read
\begin{equation}
\partial^{\alpha\dot{\alpha}}\partial_{\alpha\dot{\alpha}}\phi
=0\,,~~~~~~i\partial_{\alpha\dot{\alpha}}\lambda^{\alpha}=0\,,~~~~~f=0\,,
\label{BWG}%
\end{equation}
with analogous equations for the conjugated fields. They derive from the superspace action
\begin{equation}
S=\int[d^{4}xd^{2}\theta d^{2}\bar{\theta}]\Phi\bar{\Phi}\,. \label{BWC}%
\end{equation}
As explained in
\cite{Witten:2012bg,Castellani:2015paa}, the symbol $[d^{4}xd^{2}\theta d^{2}\bar{\theta}]$ is not a measure in the
usual sense. The form of the integral, where both $\theta$ and $\bar{\theta}$
are present, is known as non-chiral superspace integral.

There are two ways to derive the equations of motion (\ref{BWG}) from (\ref{BWC}):

1) compute the Berezin integral over $\theta$'s and $\bar{\theta}$'s to obtain
the component action:
\begin{equation}
S=\int d^{4}x\Big(\frac{1}{2}\partial^{\alpha\dot{\alpha}}\bar{\phi}%
\partial_{\alpha\dot{\alpha}}\phi+i\bar{\lambda}^{\dot{\alpha}}\partial
_{\alpha\dot{\alpha}}\lambda^{\alpha}+f\bar{f}\Big)\,, \label{BWF}%
\end{equation}
Then, derive eqs. (\ref{BWB}) by considering the variations with respect to
$\bar{\phi},\bar{\lambda}$ and $\bar{f}$.

2) vary the action with respect to the superfield $\Phi$ or $\bar{\Phi}$. This
must be done with care since they are constrained fields. First one performs a
Berezin integration over $\bar{\theta}$ leading to
\begin{equation}
S=\int[d^{4}xd^{2}\theta]\left. (\bar{D}^{2}\bar{\Phi})\Phi  \right|_{\bar\theta=0} \label{BWD}%
\end{equation}
where both $\Phi$ and $\bar{D}^{2}\bar{\Phi}$ are computed at $\bar{\theta}%
=0$. Notice that due to $D^{3}=0$ and $\bar{D}^{3}=0$ (valid in the case
$D=4$), the superfield $\bar{D}^{2}\bar{\Phi}$ is also a chiral field. The 
variation with respect to $\Phi$ gives the equations of motion (\ref{BWB}).

Likewise, one could also integrate with respect to $\theta$ to get another
version of the action
\begin{equation}
S=\int[d^{4}xd^{2}\bar{\theta}]\left.\bar{\Phi}(D^{2}\Phi) \right|_{\theta=0}
\label{BWE}%
\end{equation}
which is the anti-chiral version. Again, the equations of motion are given by
(\ref{BWB}). In (\ref{BWC}), (\ref{BWD}) or (\ref{BWE}) the
supersymmetry is manifest since they are written in terms of superfields. Any
variation of the Lagrangian under supersymmetry is a total derivative and then
the variation of the action vanishes.

In superspace, the supersymmetric transformations are implemented by 
the supersymmetry generators 
$Q_\a = \partial_{\theta^\a} + i \bar\theta^{\dot \a} \partial_{\a\dot \a}$ and 
$\bar Q_{\dot \a} = \partial_{\bar\theta^{\dot\a}} + i \theta^{\a} \partial_{\a\dot \a}$ (which commute 
with the superderivatives $D_\a$ and $\bar D_{\dot \a}$) as follows 
\begin{eqnarray}
\label{BWEAA}
\delta_{\epsilon} \Phi = (\epsilon^\a Q_\a + \bar \epsilon^{\dot\a} \bar Q_{\dot \a}) \Phi\,, ~~~~
\delta_{\epsilon} D_\a \Phi = D_\a (\delta_{\epsilon} \Phi)\,.  
\end{eqnarray}

In order to add interactions, we need to introduce the superpotential. The
superfield $\mathcal{W}(\Phi)$ is an holomorphic function of $\Phi$ (for a
renormalizable theory a polynomial of maximum degree = 3) and the
full action is written as (see \cite{GGRS}  and \cite{baggerwess})
\begin{equation}
S=\int[d^{4}xd^{2}\theta d^{2}\bar{\theta}]\Phi\bar{\Phi}+\int[d^{4}xd^{2}%
{\theta}]\mathcal{W}(\Phi)+\int[d^{4}xd^{2}\bar{\theta}]\overline
{\mathcal{W}}(\bar{\Phi})\,. \label{BWEA}%
\end{equation}
The contribution of the superpotential is automatically supersymmetric
invariant and its holomorphicity w.r.t. $\Phi$ implies the non-renormalization
properties of the WZ action. The equations of motion are computed as above, by
converting the first integral into a chiral or antichiral integral (see 
eqs. (\ref{BWD}) or (\ref{BWE})) and then varying with respect to $\Phi$
(or w.r.t. $\bar{\Phi}$) to get
\begin{equation}
D^{2}\Phi+\overline{\mathcal{W}}^{\prime}(\bar{\Phi})=0\,,~~~~~~\bar{D}%
^{2}\bar{\Phi}+\mathcal{W}^{\prime}(\Phi)=0\,. \label{BWEB}%
\end{equation}
As a consistency check observe that acting with $D_{\alpha}$ on the l.h.s.
of the first equation, both terms vanish and, similarly acting with $\bar
{D}_{\dot{\alpha}}$ on the second equation. Acting with $\bar D^2$ on the l.h.s. of 
the first equation we get 
\begin{eqnarray}
\label{BWEBA}
\bar D^2 D^2 \Phi = {\cal W}'(\Phi) \overline {\cal W}''(\bar \Phi) + \bar D_{\dot\a} \bar \Phi \bar D^{\dot\a} \bar \Phi \overline {\cal W}'''(\bar \Phi)\,,
\end{eqnarray}
which reduces to (\ref{BWB}) in absence of ${\cal W}$ and its conjugate. 

The generalization to multiple superfields $\Phi^{I}$ with $I=1, \dots, N$ is
straightforward. The superpotential $\mathcal{W}$ becomes a generic polynomial
in the superfields $\Phi^{I}$, and the kinetic term becomes a quadratic form
$\bar\Phi\Phi\rightarrow g_{\bar I J} \bar\Phi^{\bar I} \Phi^{J}$.

To couple the superfields to abelian gauge fields by minimal coupling, one
promotes to local superfield the chiral parameter $\Lambda$ of the rigid symmetry
\begin{equation}
\Phi^{I}\rightarrow e^{ie_{I}\Lambda}\Phi^{I}\,,~~~~~\bar{\Phi}^{\bar{I}%
}\rightarrow e^{-ie_{I}\overline{\Lambda}}\bar{\Phi}^{\bar{I}}\,,~~~~~
\label{BWEC}%
\end{equation}
of the action. The gauge fields are introduced by modifying the action as follows
\begin{equation}
S=\sum_{I}\int[d^{4}xd^{2}\theta d^{2}\bar{\theta}]g_{\bar{I}J}\bar{\Phi}%
^{I}e^{V}\Phi^{J}+\int[d^{4}xd^{2}{\theta}]\mathcal{W}(\Phi^{I}%
)+\int[d^{4}xd^{2}\bar{\theta}]\overline{\mathcal{W}}(\bar{\Phi}^{\bar{I}})\,.
\label{BWED}%
\end{equation}
Here $V$ is the prepotential of the gauge fields (see \cite{baggerwess} for more
details) which transforms as $V\rightarrow V+i(\Lambda-\overline{\Lambda})$.

As a final remark, one can convert the action (\ref{BWEA}) into an integral on the
complete superspace: 
\begin{equation}
S=\int[d^{4}xd^{2}\theta d^{2}\bar{\theta}]\Big(\bar{\Phi}\Phi+\mathcal{W}%
(\Phi)\bar{\theta}^{2}+\overline{\mathcal{W}}(\Phi)\theta^{2}\Big)\,.
\label{BWEF}%
\end{equation}
where we have inserted the $\theta$-terms. Integrating the second
term with respect to $\bar{\theta}$ we obtain again the chiral integral, and
likewise for the third term.

In the following, we need some algebraic relations between superderivatives.
In particular, given a superfield $\mathcal{F}_{\alpha\dot{\alpha}}%
(x,\theta,\bar{\theta})$, we need the relation
\begin{equation}
\left.  D^{2}\bar{D}^{2}\Big(\mathcal{F}_{\alpha\dot{\alpha}}\theta^{\alpha
}\bar{\theta}^{\dot{\alpha}}\Big)\right\vert_{\theta = \bar \theta =0} =\left.  D^{\alpha}\bar{D}%
^{\dot{\alpha}}\mathcal{F}_{\alpha\dot{\alpha}}\right\vert_{\theta = \bar \theta =0}
+\mathrm{total~deriv.} \label{WZI}%
\end{equation}
This implies
\begin{equation}
\int[d^{4}xd^{2}\theta d^{2}\bar{\theta}]\Big(\mathcal{F}_{\alpha\dot{\alpha}%
}\theta^{\alpha}\bar{\theta}^{\dot{\alpha}}\Big)=\int[d^{4}x]\left.
D^{\alpha}\bar{D}^{\dot{\alpha}}\mathcal{F}_{\alpha\dot{\alpha}}\right\vert_{\theta = \bar \theta =0}
\,. \label{WZL}%
\end{equation}

\subsection{Geometric WZ Action}

In the geometrical formulation, we start again from the complex scalar
superfield $\Phi$ and we impose the following condition
\begin{align}
d\Phi &  =V^{\alpha\dot{\alpha}}\partial_{\alpha\dot{\alpha}}\Phi+\psi
^{\alpha}D_{\alpha}\Phi+\bar{\psi}^{\dot{\alpha}}\bar{D}_{\dot{\alpha}}%
\Phi\label{YB}  \nonumber \\
&  =V^{\alpha\dot{\alpha}}\partial_{\alpha\dot{\alpha}}\Phi+\psi^{\alpha
}W_{\alpha}\,,
\end{align}
where $(V^{\a\dot\a}, \psi^\a, \bar\psi^{\dot\a})$ is the supervielbein (see also 
sec. 5.1). 
The differential $d$ is the usual super-differential (it is an anticommuting operator and therefore 
we assume it anticommutes with $\theta$ and $\bar \theta$ as well). Comparing the two lines,
we get 
\begin{equation}
\bar{D}_{\dot{\alpha}}\Phi=0\,,~~~~~~D_{\alpha}\Phi=W_{\alpha}\,. \label{YBA}%
\end{equation}
The new superfield $W_{\alpha}$ of (\ref{YB}) has as first component the fermion of the
supermultiplet $\lambda_{\alpha}$. Applying $d$ on the left hand side, we have
a consistency condition on $W_{\alpha}$ leading to
\begin{equation}
dW_{\alpha}=V^{\alpha\dot{\alpha}}\partial_{\alpha\dot{\alpha}}W_{\alpha
}-2i\bar{\psi}^{\dot{\alpha}}\partial_{\alpha\dot{\alpha}}\Phi+\psi_{\alpha
}F\,, \label{WZA}%
\end{equation}
where the new superfield $F$ has as first component  the auxiliary field $f$ and
$\psi_{\alpha}=\epsilon_{\alpha\beta}\psi^{\beta}$. On $W_{\alpha}$, we have
the conditions
\begin{equation}
D_{\alpha}W_{\beta}=-\epsilon_{\alpha\beta}F\,,~~~~~~\bar{D}_{\dot{\alpha}%
}W_{\alpha}=-2i\partial_{\alpha\dot{\alpha}}\Phi\,.~~~~~ \label{WZAA}%
\end{equation}
Again, applying the differential $d$, we find the differential of $F$
\begin{equation}
dF=V^{\alpha\dot{\alpha}}\partial_{\alpha\dot{\alpha}}F+2i\bar{\psi}%
^{\dot{\alpha}}\partial_{\dot{\alpha}\alpha}W^{\alpha}\,, \label{WZAB}%
\end{equation}
and the constraints 
\begin{equation}
D_{\alpha}F=0\,,~~~~~\bar{D}_{\dot{\alpha}}F=2i\partial_{\alpha\dot{\alpha}%
}W^{\alpha}\,,~~~~~F=\frac{1}{2}\epsilon^{\alpha\beta}D_{\alpha}W_{\beta
}=\frac{1}{2}\epsilon^{\alpha\beta}D_{\alpha}D_{\beta}\Phi\,,  \label{WZC}%
\end{equation}
where $\phi, \lambda^\a, f$ are the fields of the Wess-Zumino multiplet. 
It can be checked that no additional superfields are needed. The first components of
the superfields $\Phi,W^{\alpha},F$ are
\begin{equation}
\Phi=\phi+O (\theta)\,,~~~~~W_{\alpha}=\lambda_{\alpha}%
+O (\theta)\,,~~~~F=f+ O (\theta)\,. \label{WZB}%
\end{equation}
In terms of these superfields, the equations of motion are
\begin{equation}
\partial^{\alpha\dot{\alpha}}\partial_{\alpha\dot{\alpha}}\Phi
=0\,,~~~~~~\partial_{\alpha\dot{\alpha}}W^{\alpha}=0\,,~~~~~~F=0\,,
\label{WZD}%
\end{equation}
and their conjugates. These equations reduce to the spacetime
equations, by setting $\theta=\bar \theta =0$. Note that 
all components in the $\theta, \bar \theta$ expansion satisfy
the same equations, for example, by expanding the superfield at second order 
$\Phi = \phi + \theta^\a \lambda_\a + \bar \theta^{\dot\a} \bar\lambda_{\dot\a} + {\cal O}(\theta^2)$ 
we find
\begin{eqnarray}
\label{WZDA}
\partial^{\alpha\dot{\alpha}}\partial_{\alpha\dot{\alpha}}\Phi = 
\partial^{\alpha\dot{\alpha}}\partial_{\alpha\dot{\alpha}}\phi + \theta^\b 
\big(\partial^{\alpha\dot{\alpha}}\partial_{\alpha\dot{\alpha}}\lambda_\b \big) +
\bar\theta^{\dot\b} 
\big(\partial^{\alpha\dot{\alpha}}\partial_{\alpha\dot{\alpha}}\bar\lambda_{\dot\b} \big) + {\cal O}(\theta^2)
\end{eqnarray}
and $\partial^{\alpha\dot{\alpha}}\partial_{\alpha\dot{\alpha}}\lambda_\b  =0$ and 
$\partial^{\alpha\dot{\alpha}}\partial_{\alpha\dot{\alpha}}\bar\lambda_{\dot\b} =0$ 
which follow from the Dirac equations (the second eq. in (\ref{WZD}) and its conjugate) 
by acting with $\partial^{\dot\b\a}$ on $\partial_{\alpha\dot{\alpha}}W^{\alpha}=0$.

We can write the free Lagrangian $\mathcal{L}_{kin}^{(4|0)}$ for
the kinetic terms as follows
\begin{align}
\mathcal{L}_{kin}^{(4|0)}  
&  =
(V^{4}) \, (\bar{\xi}^{\alpha\dot{\alpha}}\xi_{\alpha \dot{\alpha}}+\bar{F}F)\label{WZE}\\
&  + 
(V^{3})^{\alpha\dot{\alpha}} 
\left[
(d\Phi-\psi^{\beta}W_{\beta})\bar{\xi}_{\alpha\dot{\alpha}}%
+(d\bar{\Phi}-\bar{\psi}^{\dot{\beta}}
\bar{W}_{\dot{\beta}})\xi_{\alpha\dot{\alpha}}
+a_1 (\bar{W}_{\dot{\alpha}}dW_{\alpha}+d\bar{W}_{\dot{\alpha}}W_{\alpha
})
\right]%
\nonumber\\
&  + (V^2_+)^{\a\b} \left[  a_2 (W_{\alpha}\psi_{\beta}d\bar{\Phi}) + 
a_3 (W_{\alpha}\psi_{\beta} \bar{W}^{\dot{\gamma}} \bar\psi_{\dot\gamma})
\right]
\nonumber\\
&  + (V^2_-)^{\dot\a\dot\b} \left[  a_2 (\bar W_{\dot\alpha}\bar\psi_{\dot\beta}d {\Phi}) + 
a_3 (\bar W_{\dot \alpha}\bar \psi_{\dot\beta} {W}^{{\gamma}} \psi_{\gamma})
\right]
\nonumber\\
&  + V^{\alpha\dot{\alpha}}  
\left[ a_4 (\bar{\Phi}d\Phi-d\bar{\Phi}\Phi)\psi_{\alpha}\bar{\psi}_{\dot{\alpha}%
}\right]\,.\nonumber
\end{align}
where we have adopted the following definitions (see also app. B) 
\begin{align}
V^{4}  &  =\frac{1}{4!}V_{\alpha\dot{\alpha}}\wedge V^{\dot{\alpha}\beta}\wedge V_{\beta\dot{\gamma}%
}\wedge V^{\dot{\gamma}\alpha}\,,~~~~
(V^{3})^{\alpha\dot{\alpha}}=\frac{1}{3!}V^{\alpha\dot
{\beta}}\wedge V^{\dot{\gamma}\beta}\wedge V^{\rho\dot{\alpha}}\epsilon_{\dot\beta \dot\gamma} \epsilon_{\beta\rho}    \label{WZF}\\
(V^{2}_+)^{\alpha\beta}  &  =\frac{1}{2!}V^{\alpha\dot{\beta}}\wedge V^{\dot{\beta}\beta}\epsilon_{\dot{\alpha}%
\dot{\beta}}\,,~~~~~~~~~~~~~~~
(V^{2}_-)^{\dot{\alpha}\dot{\beta}%
}=\frac{1}{2!}V^{\dot{\alpha}\alpha}\wedge V^{\beta\dot{\beta}} \epsilon_{\alpha\beta}\,,\nonumber
\end{align}
for the wedge products of the vielbeins $V^{\a\dot\a}$. 

The Lagrangian is organized in powers of $V$'s. The first line,  
proportional to the volume form $V^{4}$, contains two terms: one with 
the auxiliary fields $F$ and $\bar F$ and the other with the 
first-order-formalism field $\xi^{\alpha\dot{\alpha}}$ and its conjugate. The latter are
needed in order to write the action without using the Hodge dual operator. 
This is required for the Lagrangian to be a pure $4$-form built exclusively with fields, their
differentials and the supervielbeins. We have written all possible terms
compatible with the scaling dimensions and with the form degree. The constants
$a_1,a_2,a_3,a_4$ are fixed by requiring the closure of the Lagrangian and the
correct equations of motion. 

We have four fields $F, \xi_{\a\dot \a}, \Phi, W_\a$ and their conjugates. Therefore, 
we need four equations of motion. 

The equation of $F$ is obtained by varying $\mathcal{L}_{kin}^{(4|0)} $ 
with respect to $\bar F$. This simply gives 
\begin{eqnarray}
\label{EOMA}
(V^4) F =0
\end{eqnarray}
which is the free equation of the auxiliary field. 
The equation for the auxiliary field $\xi_{\a\dot\a}$ is 
\begin{eqnarray}
\label{EOMB}
(V^4) \xi^{\a\dot \a} + (V^3)^{\a\dot\a} 
\left( 
\psi^\b D_\b \Phi + V^{\b\dot\b} \partial_{\b\dot\b}\Phi - \psi^\b W_\b 
\right)=0 
\end{eqnarray}
which implies
\begin{eqnarray}
\label{EOMC}
W_\b = D_\b \Phi\,, ~~~~~~  \bar D_{\dot \b} \Phi =0\,, ~~~~~
\xi^{\a\dot \a} = \partial^{\a\dot \a} \Phi\,.
\end{eqnarray}
These relations identify the superfield $W_\a$ and the auxiliary field $\xi^{\a\dot\a}$ with derivatives of
$\Phi$ . In addition, the second equation establishes the chirality of the 
superfield $\Phi$. 
The equation of motion for $\Phi$ is obtained by taking the functional derivative of the 
action with respect to the superfield $\bar\Phi$. After integration by parts it becomes
\begin{eqnarray}
\label{EOMD}
&&i \left( \psi^\a (V_-^2 \bar \psi)^{\dot \a} - \bar \psi^{\dot \a} (V_+^2 \psi)^\a \right) \xi_{\a\dot\a} 
- (V^3)^{\a\dot \a} d \xi_{\a\dot\a} \nonumber \\
&& + a_2 \, \left( - 2 i \psi^{(\a} (V \bar\psi)^{\b)} \right) W_\a \psi_\b + 
a_2 (V_+^2)^{\a\b} d W_\a \psi_\b + 2 a_4 (\psi V \bar \psi) d\Phi =0\,
\end{eqnarray}
where $(\psi V \bar \psi)  = \psi^\a V^{\b\dot\b} \bar \psi^{\dot \a} \epsilon_{\a\b} \epsilon_{\dot\a\dot\b}$. 
This equation implies 
\begin{eqnarray}
\label{EOME}
\partial^{\a\dot \a} \xi_{\a\dot \a} =0 \Longrightarrow \partial^2 \Phi=0\,, ~~~~~~~
a_4 = - \frac{i}{2} a_2 \,, ~~~~ a_2=1\, \Longrightarrow a_4 = - \frac{i}{2}\,. 
\end{eqnarray}
Finally, the equation for $W_\a$ is given by 
\begin{eqnarray}
\label{EOMF}
(V^3)^{\a\dot\a} \bar \psi^{\dot \b} \xi_{\a\dot \a} - 2 a_1 (V^3)^{\alpha \dot \b} d W_\a + 
a_1 \, d (V^3)^{\a\dot\b} W_\a + 
(V^2_-)^{\dot \a\dot \b} \left(\, \bar\psi_{\dot\a} d \Phi + a_3 \, \bar \psi_{\dot \a} W \cdot \psi\right) =0 
\nonumber \\
\end{eqnarray}
(where $W \cdot \psi = W^\a \epsilon_{\a\b} \psi^\b$) yielding the equations of motion for the spinor superfield $W_\a$.  
We fix the remaining coefficients $a_1$ and $a_3$
\begin{eqnarray}
\label{EOMG}
\partial^{\a\dot \b} W_\a=0\,, ~~~~~~~ a_1 = \frac{1}{2}\,, ~~~~~ a_3 =1\,. 
\end{eqnarray}
One can check the consistency among the four equations (\ref{EOMA}), 
(\ref{EOMB}), (\ref{EOMD}), and (\ref{EOMF}).

To complete the Lagrangian we need the interaction  and the superpotential terms.
These are written as follows
\begin{align}
\mathcal{L}_{sup}^{(4|0)}  &  =\Big(\mathcal{W}^{\prime}(\Phi)F-\frac{1}%
{2}\mathcal{W}^{\prime\prime}(\Phi)W_{\alpha}W^{\alpha}\Big)(V^{4})%
+\mathcal{W}^{\prime}(\Phi)W^{\alpha}\bar{\psi}^{\dot{\alpha}}(V^{3}%
)_{\alpha\dot{\alpha}}\label{WZSUPA}\\
&  +\mathcal{W}(\Phi)\bar{\psi}^{\dot{\alpha}}\bar{\psi}^{\dot{\beta}}%
(V^{2}_-)_{\dot{\alpha}\dot{\beta}}+\mathrm{h.c.}\nonumber
\end{align}
where $\mathcal{W}(\Phi)$ is the superpotential introduced in the previous
section and $\mathcal{W}^{\prime}(\Phi),\mathcal{W}^{\prime\prime}(\Phi)$ are
the first and the second derivative of $\mathcal{W}(\Phi)$ with respect to
$\Phi$. 

The Lagrangian ${\cal L}^{(4|0)} = {\cal L}^{(4|0)}_{kin} + {\cal L}^{(4|0)}_{sup}$ 
is closed as can be verified by using the definitions of the curvatures 
$d \Phi, d W_\a, d F$ as in (\ref{YB}), (\ref{WZA}), 
(\ref{WZAB}) and the algebraic equations (\ref{EOMB}). 

\subsection{WZ Action on the Supermanifold $\mathcal{M}^{(4|4)}$}

Now we show that the action (\ref{BWEF}) 
can be obtained from the supermanifold integral 
\begin{equation}
S=\int_{\mathcal{SM}^{(4|4)}}\mathcal{L}^{(4|0)}(\Phi, W, F) \wedge{\mathbb{Y}}^{(0|4)}\,,
\label{YA}%
\end{equation}
where the Lagrangian $\mathcal{L}^{(4|0)}$ is given in the previous section. 

The PCO ${\mathbb{Y}}^{(0|4)}$ is a $(0|4)$-form which 
depends upon the superspace data. As the Lagrangian is $d$-closed, we can
shift ${\mathbb{Y}}^{(0|4)}\rightarrow{\mathbb{Y}}^{(0|4)}+d\Lambda^{(-1|4)}$
by an exact term {\it without changing the action}. The PCO's are discussed in Sec. 5.1 
(see also \cite{Castellani:2014goa,Castellani:2015paa,Castellani:2015ata,Castellani:2017ycm}).

The first PCO we consider is given by
\begin{equation}
{\mathbb{Y}}_{s.t.}^{(0|4)}=\theta^{2}\delta^{2}(\psi)\wedge\bar{\theta}%
^{2}\delta(\bar{\psi})\,, \label{WZG}%
\end{equation}
which is closed, not exact and Lorentz invariant. It is not supersymmetric, but its variation under supersymmetry is $d$-exact.  
The Dirac delta functions $\delta(\psi)$ and $\delta(\bar\psi)$ are needed to 
set $\psi$ and $\bar{\psi}$ in $\mathcal{L}^{(4|0)}$ to zero and the factor $\theta^2 \bar\theta^2$ 
sets $\theta =\bar \theta =0$. Thus the integrand (\ref{YA}) takes the form
\begin{align}
\mathcal{L}^{(4|0)}\wedge{\mathbb{Y}}_{s.t.}^{(0|4)}  &  =\Big[(\bar{\xi
}^{\alpha\dot{\alpha}}\xi_{\alpha\dot{\alpha}}+\bar{f}f)d^{4}x\label{WZH}\\
&  +\Big(d\phi\bar{\xi}^{\alpha\dot{\alpha}}+d\bar{\phi}\xi^{\alpha\dot
{\alpha}}+\frac{i}{2}(\bar{\lambda}^{\dot{\alpha}}d\lambda^{\alpha}%
+d\bar{\lambda}^{\dot{\alpha}}\lambda^{\alpha})\Big)(d^{3}x)_{\alpha
\dot{\alpha}}\nonumber\\
&  +\Big(\mathcal{W}^{\prime}(\phi)f-\frac{1}{2}\mathcal{W}^{\prime\prime
}(\phi)\lambda^{\alpha}\epsilon_{\alpha\beta}\lambda^{\beta}\Big)d^{4}%
x+\mathrm{h.c.}\Big]\theta^{2}\bar{\theta}^{2}\delta^{2}(\psi)\delta^{2}%
(\bar{\psi})\,,\nonumber
\end{align}
where $(d^{3}x)_{\a\dot{\alpha}}=dx_{\alpha\dot{\beta}}dx^{\dot{\beta}\gamma
}dx_{\gamma\dot{\alpha}}$ By solving the algebraic equations of motion for
$\xi^{\alpha\dot{\alpha}}$ and its conjugate, and using
\begin{equation}
d\phi\wedge\delta^{2}(\psi)\delta^{2}(\bar{\psi})=dx^{\alpha\dot{\alpha}%
}\partial_{\alpha\dot{\alpha}}\phi\wedge\delta^{2}(\psi)\delta^{2}(\bar{\psi
})\,,
\end{equation}
one ends up with the component Lagrangian given in (\ref{BWEA}). 
The choice of the PCO (\ref{WZG}) represents the trivial embedding of the
bosonic submanifold ${\cal M}^{4}$ into the supermanifold ${\cal M}^{(4|4)}$. 

To derive an action with 
manifest supersymmetry, we need a different PCO. That will be discussed in 
the forthcoming Section 5.1, and here we report the main result: 
\begin{eqnarray}
\label{superPCO}
{\mathbb{Y}}^{(0|4)}_{s.s.}=\Big(-4(\theta V\bar{\iota})\wedge(\bar{\theta}%
V\iota)+\theta^{2}(\iota V\wedge V\iota)+\bar{\theta}^{2}(\bar{\iota}V\wedge
V\bar{\iota})\Big)\delta^{4}(\psi) \label{NEWBA}%
\end{eqnarray}
where $\iota= \partial_\psi$ (and similar for $\bar{\iota}$). Notice that 
it still depends upon $\theta$ and $\bar \theta$. This is needed 
to produce the superspace action in the usual form. In addition, 
we notice that the first term is non-chiral and the other two are chiral and 
anti-chiral, respectively.

With this PCO, the action becomes
\begin{align}
S  &  =\int_{\mathcal{M}^{(4|4)}}\mathcal{L}^{(4|0)}\wedge{\mathbb{Y}}%
_{s.s.}^{(0|4)}\label{NEWBB}\\
&  =\int_{\mathcal{M}^{(4|4)}}\Big(\overline{W}V\psi)(\bar{\psi}%
VW)+\mathcal{W}(\Phi)(\bar{\psi}V\wedge V\bar{\psi})+\overline{\mathcal{W}%
}(\bar{\Phi})(\psi V\wedge V\psi)\Big)\wedge{\mathbb{Y}}_{s.s.}^{(0|4)}%
\nonumber\\
&  =\int_{\mathcal{M}^{(4|4)}}\Big(\overline{W}^{\dot{\alpha}}\bar{\theta
}_{\dot{\alpha}}W_{\alpha}\theta^{\alpha}+\mathcal{W}(\Phi)\bar{\theta}%
^{2}+\overline{\mathcal{W}}(\bar{\Phi})\theta^{2}\Big)V^{4}\delta^{4}%
(\psi)\nonumber\\
&  =\int[d^{4}xd^{2}\theta d^{2}\bar{\theta}]\Big(\overline{W}^{\dot{\alpha}%
}\bar{\theta}_{\dot{\alpha}}W_{\alpha}\theta^{\alpha}+\mathcal{W}(\Phi
)\bar{\theta}^{2}+\overline{\mathcal{W}}(\bar{\Phi})\theta^{2}%
\Big)\,.\nonumber
\end{align}
and, using the algebraic relations among superderivatives given in (\ref{WZI}) and (\ref{WZL}), 
recalling $W_\a = D_\a \Phi, \bar W_{\dot\a} = \bar D_{\dot\a} \bar\Phi$, 
and 
\begin{eqnarray}
\label{NEWBBA}
(D_\a \Phi )\theta^\a = D_\a ( \Phi \theta^\a) + 2 \Phi\,, 
\end{eqnarray}
and integrating by parts, 
one arrives at the usual superspace action (\ref{BWEF}). Notice that the
three pieces of the PCO ${\mathbb{Y}}^{(0|4)}_{s.s.}$ in (\ref{superPCO}) 
are essential to get the
complete action since the terms for the kinetic part and for the
superpotential have completely different algebraic structures. Notice also the
unusual form of the kinetic term which has a non-chiral structure as said above.

\vfill
\eject

\section{D=4 N=1 Integral Super Yang-Mills}

Using the same strategy, we now study the SYM action in this framework. In Sec. 3.1,
we review SYM in the superspace formulation (see \cite{baggerwess} for further
details). In Sec. 3.2 we review the geometric (rheonomic) formulation of SYM and we
discuss the equations of motion. In Sec. 3.3, we prove that both the
component action and the superspace action can be retrieved from the same
supermanifold action by changing the PCO; the same PCO given 
in (\ref{superPCO}) produces the superspace action.

\subsection{SYM superspace action}

It is convenient to adopt again a Weyl/anti-Weyl notation in order to describe
the superspace action in its most common formulation \cite{baggerwess}.
The gauge field is identified with the $(1|0)$-superconnection
\begin{equation}
A^{(1|0)}=A_{\alpha\dot{\alpha}}V^{\alpha\dot{\alpha}}+A_{\alpha}\psi^{\alpha
}+A_{\dot{\alpha}}\bar{\psi}^{\dot{\alpha}}\,, \label{SYA}%
\end{equation}
and the field strength is 
\begin{align}
\mathcal{F}  &  =dA^{(1|0)}+A^{(1|0)}\wedge A^{(1|0)}\label{SYB}\\
&  =F_{\alpha\dot{\alpha}\beta\dot{\beta}}V^{\alpha\dot{\alpha}}\wedge
V^{\beta\dot{\beta}}+F_{\alpha\dot{\alpha}\beta}V^{\alpha\dot{\alpha}}%
\wedge\psi^{\beta}+F_{\alpha\dot{\alpha}\dot{\beta}}V^{\alpha\dot{\alpha}%
}\wedge\bar{\psi}^{\dot{\beta}}\nonumber\\
&  +F_{\alpha\beta}\psi^{\alpha}\wedge\psi^{\beta}+F_{\alpha\dot{\beta}}%
\psi^{\alpha}\wedge\bar{\psi}^{\dot{\beta}}+F_{\dot{\alpha}\dot{\beta}}%
\bar{\psi}^{\dot{\alpha}}\wedge\bar{\psi}^{\dot{\beta}}\,.\nonumber
\end{align}
The superfield $A^{(1|0)}$ contains several independent components exceeding
the physical ones. Therefore, to reduce that number one needs additional
constraints. It is customary to set all spinorial field strengths to zero
\begin{equation}
F_{\alpha\beta}=0\,,~~~~~~~F_{\alpha\dot{\beta}}=0\,,~~~~~F_{\dot{\alpha}%
\dot{\beta}}=0\,. \label{SYC}%
\end{equation}
Consequently, the Bianchi identities $d\mathcal{F}+A^{(1|0)}\wedge
\mathcal{F}=0$ imply some constraints on the remaining field strengths
which can be easily solved.

The parametrizations of the curvatures are 
\begin{align}
\mathcal{F}  &  
=
F^+_{\alpha\beta}(V^2_+)^{\a\b} 
+{F}^-_{\dot{\alpha}\dot{\beta}} (V^2_-)^{\dot\a\dot\b}
+ 2 i  \, \overline{W}_{\dot{\alpha}}( V \psi)^{\dot\alpha}
+ 2 i \, W_{\alpha}(V\bar{\psi})^{{\alpha}}\label{CHIA}\\
\nabla W_{\alpha}  &  
=V^{\beta\dot{\beta}}  \nabla_{\beta\dot{\beta}}W_{\alpha}
-  (F^+\psi)_{\alpha}+{\cal D}\,\epsilon_{\a\b} \psi^{\b}\,,\nonumber\\
\nabla\overline{W}_{\dot{\alpha}}  &  
=V^{\beta\dot{\beta}}  \nabla_{\beta\dot{\beta}}\overline
{W}_{\dot{\alpha}}
- ({F}^-\bar{\psi
})_{\dot{\alpha}} - {\cal D}\, \epsilon_{\dot\a\dot\b} \bar{\psi}^{\dot{\b}}\,,\nonumber\\
\nabla {\cal D}  &  
=V^{\alpha\dot{\alpha}} \nabla_{\alpha\dot{\alpha}}{\cal D}%
-\bar{\psi}^{\dot{\alpha}} \nabla_{\alpha\dot{\alpha}}W^{\alpha}
- {\psi}^{{\alpha}}\nabla
_{\alpha\dot{\alpha}}\overline{W}^{\dot \alpha}\nonumber
\end{align}
where $(F^+\psi)_{\alpha}=F^+_{\alpha\beta}\psi^{\beta}$, 
$({F}^-\bar{\psi})_{\dot{\alpha}}={F}^-_{\dot{\alpha}\dot{\beta}}\bar{\psi}^{\dot{\b}}$, 
$W^\a = \epsilon^{\a\b} W_\b$ and $\bar W^{\dot\a} = \epsilon^{\dot\a\dot\b} \bar W_{\dot\b}$. 
The real scalar field ${\cal D}$ is an auxiliary field needed to close the algebra off-shell. 
Notice that setting ${\cal D}=0$, the last line implies the Dirac equations for $W_\a$ and 
$\bar W_{\dot \a}$: $\partial_{\a\dot \a} \bar W^{\dot \a} =0$ and $\partial_{\a\dot\a} W^\a =0$. 

The Bianchi identities for the curvatures ${\cal F}, d W_\a, d\bar W_{\dot \a}, d {\cal D}$ 
together with their parametrization (\ref{CHIA}) yield the constraints
\begin{align}
F^+_{\alpha\beta}  &  =D_{(\alpha}W_{\beta)}\,,~~~~~~\overline{D}_{\dot{\alpha}%
}W^{\beta}=0\,,~~~~~~{\cal D}=D_{\alpha}W^{\alpha}\,,\label{CHIB}\\
{F}^-_{\dot{\alpha}\dot{\beta}}  &  =\overline{D}_{(\dot{\alpha}%
}\overline{W}_{\dot{\beta})}\,,~~~~~~D_{\alpha}\overline{W}^{\dot{\beta}%
}=0\,,~~~~~~{\cal D}=\overline{D}_{\dot{\alpha}}\overline{W}^{\dot{\alpha}%
}\,,\nonumber
\end{align}
and 
\begin{eqnarray}
\label{CRIBBIOA}
D_\rho F^-_{\dot \a \dot \b} = - 2i \, \partial_{\rho (\dot \a} \overline W_{\dot \b)}\,, ~~~~~~~
\overline {D}_{\dot\rho} F^-_{\dot \a \dot \b} = - 2i \, \epsilon_{\dot \rho (\dot \a} 
\partial_{\dot \b) \rho} W^{\rho}\,, ~~~~~~~\nonumber \\
\overline{D}_{\dot\rho} F^+_{\a \b} = - 2i \, \partial_{\dot\rho (\a} W_{\b)}\,, ~~~~~~~
{D}_{\rho} F^+_{\a \b} = - 2i \, \epsilon_{\rho (\a} 
\partial_{\b) \dot\rho} \overline W^{\dot\rho}\,, ~~~~~~~
\end{eqnarray}
The latter can be verified by using (\ref{CHIB}) together with the 
algebra of superderivatives and with the Schouten identities 
$\epsilon_{\rho\a} \epsilon^{\tau \sigma} = 
(\delta^\tau_\a \delta^\sigma_\rho - \delta^\tau_\rho \delta^\sigma_\a)$ and 
$\epsilon_{\dot\rho\dot\a} \epsilon^{\dot\tau \dot\sigma} = 
(\delta^{\dot\tau}_{\dot\a} \delta^{\dot\sigma}_{\dot\rho} - 
\delta^{\dot\tau}_{\dot\rho} \delta^{\dot\sigma}_{\dot\a})$.

The second equation of the first line of (\ref{CHIB}) implies that the superfield $W^{\alpha}$
is chiral and therefore can be decomposed as follows
\begin{equation}
W_{\alpha}=\lambda_{\alpha}+(f_{\alpha\beta}+\epsilon_{\alpha\beta}\hat
{\cal D})\theta^{\beta}+\frac{1}{2}\partial_{\alpha\dot{\alpha}}\overline{\lambda
}^{\dot{\alpha}}\theta^{2} \label{CHIC}%
\end{equation}
where $\lambda_{\alpha}(x),\overline{\lambda}_{\dot{\alpha}}(x)$ are the
Weyl/anti-Weyl components of the gaugino, $f_{\alpha\beta}(x),f_{\dot
{\alpha}\dot{\beta}}(x)$ are the self-dual and anti-self dual part of the
Maxwell tensor and $\hat{\cal D}$ is the real auxiliary field (the first component
of ${\cal D}=\hat{\cal D}(x)+O(\theta)$).

In terms of these fields the superspace action can be written as
\begin{equation}
S=\int[d^{4}xd^{2}\theta]W^{\alpha}W_{\alpha}+\int [d^{4}x d^{2}\bar{\theta}]\bar
{W}_{\dot{\alpha}}\bar{W}^{\dot{\alpha}} \label{CHID}%
\end{equation}
separating the chiral and the antichiral part. Again, as in the WZ case, we
can rewrite the action as an integral on the full superspace (non-chiral
integral) as follows
\begin{equation}
S=\int[d^{4}x d^{2}\theta d^{2}\bar{\theta}]\Big(W^{\alpha}W_{\alpha}\bar
{\theta}^{2}+\bar{W}_{\dot{\alpha}}\bar{W}^{\dot{\alpha}}\theta^{2}%
\Big) \label{CHIE}%
\end{equation}
where the powers of $\theta$ and $\bar{\theta}$ are needed to
reproduce the correct action. The last equation (\ref{CHIE}) will be useful for the 
comparison with the supermanifold approach. 

\subsection{Geometric SYM action}

%

Following the method described in the  book \cite{Castellani}, based on scaling dimensions of 
the fields, form degree, Lorentz invariance and gauge invariance,  
the geometric (rheonomic) Lagrangian for N=1 super Yang-Mills is found to be 
\begin{align}
\mathcal{L}^{(4|0)}  
&  = 
\mathrm{Tr}(\mathcal{F} F^+_{\a\b}){}_{\wedge} ({V}^2_+)^{\a\b} + 
\mathrm{Tr}(\mathcal{F} F^-_{\dot\a\dot\b}){}_{\wedge} (V^2_-)^{\dot\a\dot\b} 
\nonumber \\
&- 
\mathrm{Tr}(F^+_{\a\b} F^{+ \a\b}  + F^-_{\dot\a\dot\b} F^{- \dot\a\dot\b} + \frac12 {\cal D}^2) (V^4)
\label{SYMAquattro}\nonumber \\
&  - \frac12 \, \mathrm{Tr}(\bar{W}_{\dot\a}\nabla W_\a +  \nabla \bar{W}_{\dot\a} W_\a){}_{\wedge} (V^3)^{\dot \a \a}
\nonumber\\
&  - 4i  \Big( 
\mathrm{Tr}(F^+_{\a\b} \bar{W}_{\dot \sigma}) (V^3)^{\a\dot\sigma} {}_{\wedge} \psi^{\beta} - 
\mathrm{Tr}(F^-_{\dot\a\dot\b} \bar{W}^{\dot \beta}) (V^3 \psi)^{\dot\alpha} 
\nonumber \\
&
- \mathrm{Tr}(F^+_{\a\b} {W}^{\beta}) (V^3 \bar\psi)^{\a} 
+ \mathrm{Tr}(F^-_{\dot\a\dot\b} {W}_{\sigma}) (V^3)^{\dot\a\sigma} {}_{\wedge} \bar\psi^{\dot\b} \Big) + 
\nonumber \\
& + 
2 i \,\Big(
\mathrm{Tr}(\mathcal{F}\bar{W}_{\dot \a}){}_{\wedge} (V \psi)^{\dot \a} +
\mathrm{Tr}(\mathcal{F} {W}_{\a}){}_{\wedge} (V \bar\psi)^\a \Big)
\nonumber\\
& 
+ 2 
\Big(
\mathrm{Tr}(W_{\a} W_{\b})\epsilon^{\a\b} (\bar\psi V^2_- \bar\psi) +
\mathrm{Tr}(\bar W_{\dot\a} \bar W_{\dot\b})\epsilon^{\dot\a\dot\b} (\psi V^2_+ \psi) 
\Big)
\end{align}
The Lagrangian is closed, by using the parametrization of curvatures (\ref{CHIA}) and the 
algebraic equation for $F_{\a\b}^+$ and for $F^{-}_{\dot\a\dot\b}$. The closure of ${\cal L}^{(4|0)}$ implies also the supersymmetry invariance 
of the action since ${\ell}_\epsilon {\cal L}^{(4|0)} = d \iota_\epsilon  {\cal L}^{(4|0)}$. 

The first three lines contain those terms which reduce to the component action by using the
simplest PCO
\begin{equation}
{\mathbb{Y}}^{(0|4)}_{s.t.}=\theta^{2} \bar\theta^2\delta^{2}(\psi) \delta^{2}(\bar\psi)\,. \label{SYMBquattro}%
\end{equation}
The action is
\begin{eqnarray}
\label{SYYA}
S = \int_{{\cal M}^{(4|4)}} \mathcal{L}^{(4|0)}(A, F^\pm, W, \bar W)  \wedge {\mathbb{Y}}^{(0|4)}_{s.t.}\,. 
\end{eqnarray}
The Dirac delta's for $\psi$ and $\bar\psi$ set the last four lines to zero, whereas
the factor $\theta^{2} \bar\theta^2$ extracts the lowest components of the superfields
$\mathcal{F},W_\a$ and $\bar W_{\dot\a}$. These coincide with the curvature of the gauge field (after
using the algebraic equations of motion for $F_{\a\b}^+, F_{\dot\a\dot\b}^-$) 
and with the gauginos,
respectively.

\subsection{SYM Action on the Supermanifold $\mathcal{M}^{(4|4)}$}

The way to get the superspace action is to consider the following supermanifold integral
\begin{equation}
S_{SYM}=\int_{\mathcal{M}^{(4|4)}}\mathcal{L}^{(4|0)}\wedge{\mathbb{Y}%
}^{(0|4)}_{s.s.} \label{SYMCquattro}%
\end{equation}
where the integral is extended to the full supermanifold ${\cal M}^{(4|4)}$. Now, in order to
reproduce the superspace action, we use the real PCO discussed in Sec. 2.3 (see also Sec 5.1 for the
computational details). 
\begin{equation}
{\mathbb{Y}}^{(0|4)}=\Big(-4(\theta V\bar{\iota})\wedge(\bar{\theta}%
V\iota)+\theta^{2}(\iota V\wedge V\iota)+\bar{\theta}^{2}(\bar{\iota}V\wedge
V\bar{\iota})\Big)\delta^{4}(\psi) \label{PCOA}%
\end{equation}


The last two terms in the Lagrangian (\ref{SYMAquattro}) can be rewritten as
follows
\begin{align}
& 2\int  
\Big(
\mathrm{Tr}(W_{\a} W_{\b})\epsilon^{\a\b} (\bar\psi V^2_- \bar\psi) +
\mathrm{Tr}(\bar W_{\dot\a} \bar W_{\dot\b})\epsilon^{\dot\a\dot\b} (\psi V^2_+ \psi) 
\Big)
 \wedge
{\mathbb{Y}}^{(0|4)}=\label{CHID1}\\
&  \int\left(  W_{\rho}W^{\rho}\omega^{(4|2)}\bar{\theta}^{2}\delta^{2}%
(\bar{\psi})+\mathrm{h.c.}\right)  =\int [d^4x d^2\theta] W^{\alpha}W_{\alpha}%
+\int [d^4x d^2 \bar\theta] \overline{W}_{\dot{\alpha}}\overline{W}^{\dot{\alpha}%
}\nonumber
\end{align}
where $\omega^{(4|2)} = V^4 \delta^2(\psi)$ is the chiral volume form discussed in Sec. 5.3. 
The last two integrals are computed with respect to the chiral superspaces $(x,\theta)$ and
$(x,\bar{\theta})$. The final answer coincides with the usual superspace Lagrangian. Notice 
that there is no other contribution from the complicated action (\ref{SYMAquattro}) because 
of the power of $V$'s and the derivatives of delta functions. 

\section{D=4 N=2 Integral SYM}

\subsection{N=2 Vector Superfields}

To discuss the N=2 case, we consider the simplest case, namely the N=2 vector
multiplet. This contains $4 \oplus4$ on-shell degrees of freedom. The
superspace is described by the coordinates $(x^{a}, \theta^{\alpha}_{A},
\bar\theta_{A}^{\dot\alpha})$ with $A=1,2$.

These degrees of freedom are easily understood in terms of N=1 superfields:
one chiral superfield $\Phi$ and one real superfield $V$ (better expressed in
terms of the chiral superfield $W^{\alpha}$). The off-shell degrees of freedom
are $3$ bosonic d.o.f. for the gauge field (with one gauge degree of freedom),
$1$ d.o.f. for the auxiliary field ${\cal D}$, a complex scalar $\phi$ and the
complex auxiliary field $F$; on the other side, there are $8$ fermions for the
$N=2$ gaugino.

We define a N=2 chiral superfield as a complex scalar superfield $\Psi$
constrained by the conditions
\begin{equation}
\bar{D}_{\dot{\alpha},A}\Psi=0\,,~~~~~~A=1,2 \label{enne2A}%
\end{equation}
where $D_{\dot{\alpha}A}$ is the superderivative with the algebra
$\{D_{\alpha}^{A},D_{\beta}^{B}\}=0$ and $\{D_{\alpha}^{A},\bar{D}%
_{B\dot{\beta}}\}=2i\delta_{~B}^{A}\gamma_{\alpha\dot{\beta}}^{a}\partial_{a}%
$. Solving the constraints, we get the expression
\begin{equation}
\Psi(x,\theta_{A})=\Phi(x,\theta_{1})+W_{\alpha}(x,\theta_{1})\theta
_{2}^{\alpha}+F(x,\theta_{1})\,(\theta_{2})^{2} \label{enne2B}%
\end{equation}
where $F$ is related to the complex conjugate of $\Phi$ and of $W^{\alpha}$
(see \cite{Grimm:1977xp,AlvarezGaume:1996mv}).

 In (\ref{enne2B}), we expanded the superfield $\Psi$ in
terms of $\theta_{2}^{\alpha}$. The components $\Phi,W^{\alpha},F$ are
superfields depending on $(x^{a},\theta_{1}^{\alpha})$. The action for the vector superfield $\Psi$ 
reads
\begin{equation}
S=\mathrm{Im}\frac{1}{2}\int [d^{4}xd^{2}\theta_1 d^2 \theta_2] \,\Psi^{2} \label{enne2C}%
\end{equation}
Performing the Berezin integral over $\theta_{2}^{a}$ produces the action of
N=1 superfield $W^{\alpha}$ coupled to a chiral superfield $\Phi$.

Let us move to the rheonomic action. We should consider the rheonomic
parametrization. The first equation is
\begin{equation}
d\Psi=\partial_{a}\Psi V^{a}+\lambda_{\alpha}^{A}\psi_{A}^{\alpha}\,,
\label{enne2D}%
\end{equation}
where we have denoted by $\lambda_{\alpha}^{A}=D_{\alpha}^{A}\Psi$ the
gauginos. In the same way we define the Maxwell tensor $F^+, F^-$ and the scalar $\Phi = A + i B$
\begin{align}
&  F^+_{\alpha\beta}=\epsilon_{AB}D_{(\alpha}^{A}\lambda_{\beta)}^{B}%
\,,~~~~~~~~~\overline{F}^-_{\dot{\alpha}\dot{\beta}}=\epsilon^{AB}\overline
{D}_{A}^{(\dot{\alpha}}\overline{\lambda}_{B}^{\dot{\beta})}\,, \label{enne2F}%
\\
&  \Phi=\epsilon_{AB}\epsilon^{\alpha\beta}D_{\alpha}^{A}\lambda_{\beta}%
^{B}\,,~~~~~~~~~\overline{\Phi}=\epsilon^{AB}\epsilon_{\dot{\alpha}\dot{\beta
}}\overline{D}_{A}^{\dot{\alpha}}\overline{\lambda}_{B}^{\dot{\beta}}\nonumber
\end{align}
In terms of those fields, the rheonomic action is given in the book \cite{Castellani} (in eq. II.9.34). 
Here we reproduce only the relevant terms  
\begin{equation}
\mathcal{L}_{rheo}^{(4|0)}=\dots+\frac{i}{4}(A^{2}-B^{2})\epsilon^{AB}%
\epsilon^{CD}\overline{\psi}_{A}\psi_{B}\overline{\psi}_{C}\gamma_{5}\psi
_{D}+\frac{1}{4}A\,B\epsilon^{AB}\epsilon^{CD}\overline{\psi}_{A}\psi
_{B}\overline{\psi}_{C}\psi_{D}+\dots\label{enne2G}%
\end{equation}
where the ellipsis stand for other terms of the action which do not contribute. The superfields $A$ and $B$ are the real and
imaginary part of the chiral superfield $\Phi$. We selected those terms of the
action which contain four gravitinos $\psi_{A}$. All other terms contain at
least one power of $V^{a}$. 

Now we study the PCO. As discussed in the above sections, we have the
simplest PCO 
\begin{equation}
{\mathbb{Y}}^{(0|8)}=\theta^{8}\delta^{8}(\psi) \label{enne2H}%
\end{equation}
where
\begin{equation}
\theta^{8}=(\epsilon^{AC}\epsilon^{BD}\epsilon_{\alpha\beta}\epsilon
_{\gamma\delta}\theta_{A}^{\alpha}\theta_{B}^{\beta}\theta_{C}^{\gamma}%
\theta_{D}^{\delta})(\epsilon^{\dot{\alpha}\dot{\beta}}\epsilon^{\dot{\gamma
}\dot{\delta}}\epsilon_{AC}\epsilon_{BD}\bar{\theta}_{\dot{\alpha}}^{A}%
\bar{\theta}_{\dot{\beta}}^{B}\bar{\theta}_{\dot{\gamma}}^{C}\bar{\theta
}_{\dot{\delta}}^{D})
\end{equation}
and equivalently for $\delta^{8}(\psi)$. The PCO is closed and not
exact. Computing the action
\begin{equation}
S=\int_{\mathcal{M}^{(4|8)}}\mathcal{L}_{rheo}^{(4|0)}\wedge{\mathbb{Y}%
}^{(0|8)} \label{enne2I}%
\end{equation}
we get the component action for $N=2$ SYM in $d=4$. Since the Lagrangian  
$\mathcal{L}_{rheo}^{(4|0)}$ is
closed we can change the PCO at will (in the same cohomology class). In particular, we can
choose a supersymmetric PCO. For this we notice that we can construct such an
operator by multiplying two PCO's of the $N=1$ type given in sec. (5.3):
\begin{equation}
{\mathbb{Y}}_{A}^{(0|4)}=V^{a}\wedge V^{b}(\bar{\theta}_{A}^{2}\iota_{A}%
\gamma_{ab}\iota_{A}+\mathrm{h.c.})\delta^{4}(\psi_{A})\,,~~~~~~~A=1,2
\label{enne2J}%
\end{equation}
where $\iota_{A \a} = \partial/ \partial \psi^{\a A}$ and we obtain
\begin{align}
{\mathbb{Y}}^{(0|8)}  &  =V^{a}\wedge V^{b}(\bar{\theta}_{1}^{2}\iota
_{1}\gamma_{ab}\iota_{1}+\mathrm{h.c.})\wedge V^{c}\wedge V^{d}(\bar{\theta
}_{2}^{2}\iota_{2}\gamma_{cd}\iota_{2}+\mathrm{h.c.})\delta^{8}(\psi
)\label{enne2K}\\
&  =V^{4}\epsilon^{abcd}\left(  \bar{\theta}^{4}\iota_{1}\gamma_{ab}\iota
_{1}\iota_{2}\gamma_{cd}\iota_{2}+\mathrm{h.c.}\right)  \delta^{8}%
(\psi)\nonumber
\end{align}
which is closed and not exact. Notice that closure is easily verified by using
the MC equations $dV^{a}=\bar{\psi}^{A}\gamma^{a}\psi_{A}$. The presence of
the factor $\bar{\theta}^{4}$ is essential for the non-exactness. The other
terms are needed to have a real PCO.

The main issue is the overall factor $V^{4}$. This is due to the two factors
$V^{a}$ in the factorized PCO's ${\mathbb{Y}}^{(0|4)}_{A}$ and to their
anti-symmetrization. That factor is essential to provide the bosonic part of
the volume integral form. On the other side, the four derivatives
$\iota_{\alpha}^{A}$ must act on four gravitino terms in the action.
Thus the four-gravitino terms of the action (\ref{enne2I}) are selected, giving a
term proportional to the scalar $(A + i B)^{2} = (A^{2} - B^{2}) + 2 i AB$. In
addition, the PCO selects the chiral part of the superfields leading to the
correct action (\ref{enne2C}).


\section{The geometry of D=4 N=1 Supermanifolds}

The integral forms are the crucial ingredients to define a geometric
integration theory for supermanifolds inheriting all the good properties of
integration theory in conventional (purely bosonic)
geometry. In this section we briefly describe the notations and the most
relevant definitions (see \cite{Witten:2012bg} and also
\cite{Castellani:2015paa,Castellani:2014goa,Castellani:2015ata,Castellani:2017ycm}). We introduce
the complexes of superforms, of integral forms and of pseudo-forms. These
complexes are represented in the figure 1 below. Horizontally the operator
is the usual odd differential, vertically (up and down)  the
\textit{picture changing operators} (PCO's) map cohomology classes into
cohomology classes. The PCO's are not coboundary operators, so fig.1 does not
represent a double complex. The complexes are filtered by two numbers (the
form number and the picture number) as described below.

The present section is organized as follows: 1) we first review some of the
properties of the complex of superforms and the differential operators acting
on it, 2) we discuss the properties of forms under Lorentz and linear
transformations, 3) we discuss the space of superfields and of the volume
forms, 4) we construct a few new cohomology classes needed for applications,
5) we build the PCO of type $\mathbb{Y}$ (raising the picture number)  with manifest supersymmetry, 6) we check that by
consistency the action of the PCO of type $\mathbb{Z}$ (lowering the picture number) indeed 
maps cohomology into cohomology. In Sec. 5.7 we rederive, in the integral form framework,
 two well-known theorems for superspace field theories.

\subsection{Flat D=4 N=1 Integral Superspace}

Let us first discuss the $D=4$ $N=1$ supermanifold $\mathcal{M}^{(4|4)}$.
Locally it is described in terms of the coordinates $(x^{\alpha\dot{\alpha}%
},\theta^{a},\bar{\theta}^{\dot{\alpha}}$ with $\alpha,\dot{\alpha}=1,2$) of
the superspace ${\mathbb{R}}^{(4|4)}$. We recall that $x^{a}=x^{\alpha
\dot{\alpha}}\gamma_{\alpha\dot{\alpha}}^{a}$ ($a=1...4,$ see {Appendix A}
for details on the relations between vectorial and chiral notations). We will
use the notation $(4|4)$ to denote quantities in the real representation, and
the notations $(4|2,0)$ and $(4|0,2)$ for chiral (or anti-chiral) quantities.

Let us fix our conventions. We define the flat supervielbeins
\begin{equation}
V^{\alpha\dot{\alpha}}=dx^{\alpha\dot{\alpha}}+i(\theta^{\alpha}d\bar{\theta
}^{\dot{\alpha}}+d\theta^{\alpha}\bar{\theta}^{\dot{\alpha}})\,,~~~~~\psi
^{\alpha}=d\theta^{\alpha}\,,~~~~\bar{\psi}^{\dot{\alpha}}=d\bar{\theta}%
^{\dot{\alpha}}\,,
\end{equation}
which satisfy
\begin{equation}
dV^{\alpha\dot{\alpha}}=2\,i\psi^{\alpha}\wedge\bar{\psi}^{\dot{\alpha}%
}\,,~~~~d\psi^{\alpha}=0\,,~~~~d\bar{\psi}^{\dot{\alpha}}=0\,.
\end{equation}
We also denote the derivatives as follows
\begin{equation}
\partial_{\alpha\dot{\alpha}}\,,~~~~~D_{\alpha}=\frac{\partial}{\partial
\theta^{\alpha}}-i\bar{\theta}^{\dot{\alpha}}\partial_{\alpha\dot{\alpha}%
}\,,~~~~\bar{D}_{\dot{\alpha}}=\frac{\partial}{\partial\bar{\theta}%
^{\dot{\alpha}}}-i\theta^{\alpha}\partial_{\alpha\dot{\alpha}}\,,~~~~
\end{equation}
with the commutation relations
\begin{equation}
\{D_{\alpha},D_{\beta}\}=0\,,~~~~~\{\bar{D}_{\dot{\alpha}},\bar{D}_{\dot
{\beta}}\}=0\,,~~~~~\{D_{\alpha},\bar{D}_{\dot{\alpha}}\}=-2i\partial
_{\alpha\dot{\alpha}}\,,~~~~~
\end{equation}
while $\partial_{\alpha\dot{\alpha}}$ commutes with the other differential
operators. We introduce the contraction operators
\begin{equation}
\iota_{\alpha\dot{\alpha}}=\iota_{\partial_{\alpha\dot{\alpha}}}%
\,,~~~~~\iota_{\alpha}\equiv\iota_{D_{\alpha}}=\iota_{\partial_{\alpha}}%
-i\bar{\theta}^{\dot{\alpha}}\iota_{\alpha\dot{\alpha}}\,,~~~~~\bar{\iota
}_{\dot{\alpha}}=\iota_{\bar{D}_{\dot{\alpha}}}=\bar{\iota}_{\partial
_{\dot{\alpha}}}-i\theta^{\alpha}\iota_{\alpha\dot{\alpha}}\,,
\end{equation}
where $\iota_{\partial_{\alpha}}\equiv\frac{\partial}{\partial\psi^{\alpha}}$
and $\iota_{\partial_{\dot{\alpha}}}=\frac{\partial}{\partial\bar{\psi}%
^{\dot{\alpha}}}$. The following relations hold:
\begin{align}
&  \iota_{\alpha\dot{\alpha}}V^{\beta\dot{\beta}}=\delta_{\alpha}^{~\beta
}\delta_{\dot{\alpha}}^{~\dot{\beta}}\,,~~~~~\iota_{\alpha}\psi^{\beta}%
=\delta_{\alpha}^{\beta}\,,~~~~~\iota_{\dot{\alpha}}\bar{\psi}^{\dot{\beta}%
}=\delta_{\dot{\alpha}}^{~\dot{\beta}}\,,\label{DEF_F}\\
&  \iota_{\alpha\dot{\alpha}}\psi^{\beta}=\iota_{\alpha\dot{\alpha}}\bar{\psi
}^{\dot{\beta}}=0\,,~~~~\iota_{\alpha}V^{\beta\dot{\beta}}=\iota_{\alpha}%
\bar{\psi}^{\dot{\beta}}=0\,,~~~~\bar{\iota}_{\dot{\alpha}}V^{\beta\dot{\beta
}}=\bar{\iota}_{\dot{\alpha}}\psi^{\beta}=0\,,~~~~\nonumber
\end{align}
The contraction operator $\iota_{\alpha\dot{\alpha}}$ is an odd
differential operator, while $\iota_{\alpha}$ and $\bar{\iota}_{\dot{\alpha}}$
are even. Their (anti)commutation relations are all vanishing.

\begin{figure}[h]
\begin{center}%
\begin{tabular}
[c]%
{@{\hskip1.5mm}c@{\hskip1.5mm}c@{\hskip1.5mm}c@{\hskip1.5mm}c@{\hskip1.5mm}c@{\hskip1.5mm}c@{\hskip1.5mm}c@{\hskip1.5mm}c@{\hskip1.5mm}c@{\hskip1.5mm}c}
&  &  &  &  &  &  &  &  & \\[-0.34cm]%
$\Omega^{(0|0)}$ & $\overset{d}{\longrightarrow}$ & $\Omega^{(1|0)}$ &
$\overset{d}{\longrightarrow}$ & $\Omega^{(2|0)}$ & $\overset{d}%
{\longrightarrow}$ & $\Omega^{(3|0)}$ & $\overset{d}{\longrightarrow}$ &
$\Omega^{(4|0)}$ & \\[0.05cm]%
{\large {$\updownarrow$}} &  & {\large {$\updownarrow$}} &  &
{\large {$\updownarrow$}} &  & {\large {$\updownarrow$}} &  &
{\large {$\updownarrow$}} & \\[0.08cm]%
$\Omega^{(0|1)}$ & $\overset{d}{\longrightarrow}$ & $\Omega^{(1|1)}$ &
$\overset{d}{\longrightarrow}$ & $\Omega^{(2|1)}$ & $\overset{d}%
{\longrightarrow}$ & $\Omega^{(3|0)}$ & $\overset{d}{\longrightarrow}$ &
$\Omega^{(4|0)}$ & \\[0.05cm]%
{\large {$\updownarrow$}} &  & {\large {$\updownarrow$}} &  &
{\large {$\updownarrow$}} &  & {\large {$\updownarrow$}} &  &
{\large {$\updownarrow$}} & \\[0.08cm]%
$\Omega^{(0|2)}$ & $\overset{d}{\longrightarrow}$ & $\Omega^{(1|2)}$ &
$\overset{d}{\longrightarrow}$ & $\Omega^{(2|2)}$ & $\overset{d}%
{\longrightarrow}$ & $\Omega^{(3|2)}$ & $\overset{d}{\longrightarrow}$ &
$\Omega^{(4|2)}$ & \\[0.05cm]%
{\large {$\updownarrow$}} &  & {\large {$\updownarrow$}} &  &
{\large {$\updownarrow$}} &  & {\large {$\updownarrow$}} &  &
{\large {$\updownarrow$}} & \\[0.08cm]%
$\Omega^{(0|3)}$ & $\overset{d}{\longrightarrow}$ & $\Omega^{(1|3)}$ &
$\overset{d}{\longrightarrow}$ & $\Omega^{(2|3)}$ & $\overset{d}%
{\longrightarrow}$ & $\Omega^{(3|3)}$ & $\overset{d}{\longrightarrow}$ &
$\Omega^{(4|3)}$ & \\[0.05cm]%
{\large {$\updownarrow$}} &  & {\large {$\updownarrow$}} &  &
{\large {$\updownarrow$}} &  & {\large {$\updownarrow$}} &  &
{\large {$\updownarrow$}} & \\[0.08cm]%
$\Omega^{(0|4)}$ & $\overset{d}{\longrightarrow}$ & $\Omega^{(1|4)}$ &
$\overset{d}{\longrightarrow}$ & $\Omega^{(2|4)}$ & $\overset{d}%
{\longrightarrow}$ & $\Omega^{(3|4)}$ & $\overset{d}{\longrightarrow}$ &
$\Omega^{(4|4)}$ & \\[0.05cm]
&  &  &  &  &  &  &  &  &
\end{tabular}
\end{center}
\par
\vskip -0.8cm\caption{\textrm{{\small The complex of pseudoforms for the
supermanifold $\mathcal{M}^{(4|4)}$.} }}%
\label{twocomplex}%
\end{figure}

The first row in Fig. 1 is the complex of superforms and the last one is the complex of
integral forms (the pseudoforms of maximal picture). The differential $d$ is
the usual odd differential. Along the vertical line (up and down), the picture
changing operators (PCO's) act by increasing or decreasing the picture (i.e.the
number of delta forms).

We denote by $\Omega$ the space of all pseudoforms. It is filtered by two
integers numbers $p$ and $q$:
\begin{equation}
\Omega=\bigoplus_{p,q}\Omega^{(p|q)}\left(  \mathcal{M}^{(4|4)}\right)
\end{equation}
where $q$ denotes the picture number and $p$ is the form number. The picture
$q$ ranges between $0\leq q\leq4$. The range of values for $p$ depends on $q$.
At picture zero ($q=0)$, we have the space of superforms $\Omega^{(p|0)}.$ A
generic element $\omega^{(p|0)}$ is given by:
\begin{equation}
\omega^{(p|0)}=\sum_{r,s,t,r+s+t=p}\omega_{{\small [a_{1}\dots a_{r}%
](\alpha_{1}\dots\alpha_{s})(\dot{\alpha}_{1}\dots\dot{\alpha}_{t})}}V^{a_{1}%
}\dots V^{a_{r}}\psi^{\alpha_{1}}\dots\psi^{\alpha_{s}}\bar{\psi}^{\dot
{\alpha}_{1}}\dots\bar{\psi}^{\dot{\alpha}_{t}}%
\end{equation}
where the coefficients $\omega_{{\small [a_{1}\dots a_{r}](\alpha_{1}%
\dots\alpha_{s})(\dot{\alpha}_{1}\dots\dot{\alpha}_{t})}}(x,\theta,\bar
{\theta})$ are superfields. There is no upper bound in the number of $\psi$'s
and $\bar{\psi}$'s, therefore $p\geq0$ for $q=0$. However, it will be seen
that there are no nontrivial cohomology classes for $p>4$. The total form
number is
\begin{equation}
p=r+s+t\,.
\end{equation}

At maximal picture we have the space of the integral forms $\Omega^{(p|4)}.$ A
generic element $\omega^{(p|4)}$ is given by:
\begin{equation}
\omega^{(p|4)}=\sum_{r}\sum_{\beta_{1}\beta_{2}}\sum_{\gamma_{1}\gamma_{2}%
}\omega_{{\small [a_{1}\dots a_{r}]}}V^{a_{1}}\dots V^{a_{r}}\delta
^{(\beta_{1})}(\psi^{1})\delta^{(\beta_{2})}(\psi^{2})\delta^{(\gamma_{1}%
)}(\bar{\psi}^{\dot{1}})\delta^{(\gamma_{2})}(\bar{\psi}^{\dot{2}})
\end{equation}
where $\delta^{(\beta_{1})}(\psi^{1})=(\iota_{1})^{\beta_{1}}\delta(\psi
^{1})=\frac{\partial^{\beta_{1}}}{\partial(\psi^{1})^{\beta_{1}}}\delta
(\psi^{1})$ denotes the $\beta_{1}-$ th derivative of $\delta(\psi^{1})$ with
respect to its argument (and analogously for the other terms in the monomial).
The derivatives of the delta's carry negative form degree and therefore the
total form number of $\omega^{(p|4)}$ is
\begin{equation}
p=r-(\beta_{1}+\beta_{2}+\gamma_{1}+\gamma_{2})\,.
\end{equation}
Thus the complex of integral forms is bounded from above, but it
is unbounded from below. Notice that $\delta(\psi)$ and $\delta(\bar{\psi})$
carry zero form degree and that $\psi^{1}\iota_{1}\delta(\psi^{1}%
)=-\delta(\psi^{1})$ (and analogously for the other terms).

For $p>4$, $\Omega^{(p|4)}=0$, but we can have any negative-degree integral form in the spaces
$\Omega^{(-p|4)}$ with $p>0$. It is important to notice that each space
$\Omega^{(p|4)}$ for any $p$ is finitely generated and that its dimension
increases when the form degree decreases. This parallels the case
of superforms whose complex is also finitely generated with a dimension that
increases  with higher $\psi$ and $\bar{\psi}$ powers. That is the basis
for establishing the Hodge dual correspondence between the two complexes
\begin{equation}
\star:\Omega^{(p|0)}(\mathcal{M})\longrightarrow\Omega^{(4-p|4)}(\mathcal{M})
\end{equation}
as discussed in \cite{Castellani:2015ata} and \cite{LMP}.

Finally, we have the spaces of pseudoforms with $0<q<4$. Each space
$\Omega^{(p|q)}$ is not finitely generated and these complexes are unbounded
from above and from below. Since there are no nontrivial cohomology classes
in $\Omega^{(p|q)}$ with $p>4$ and $p<0$ (as discussed for example in
\cite{Catenacci:2010cs}), we restrict our analysis to the square box formed by
the complexes $\Omega^{(p|q)}$ with $0\leq q\leq4$ and $0\leq p\leq4$. Note that even
for pseudoforms there is a Hodge duality operator
\begin{equation}
\star:\Omega^{(p|q)}\longrightarrow\Omega^{(4-p|4-q)}%
\end{equation}

We consider now some operators. The odd differential $d$ acts
horizontally
\begin{equation}
d:\Omega^{(p|q)}\longrightarrow\Omega^{(p+1|q)}%
\end{equation}
increasing the form degree and leaving unmodified the picture number. We
have already introduced the contraction operators $\iota_{a},\iota_{\alpha
},\iota_{\dot{\alpha}}$ and consequently the Lie derivatives ${\cal L}_a = i_a d + d i_a$ etc.
The $d-$cohomology is well-defined in the present framework and we denote by
$H_{d}(\Omega^{(p|q)})$ de Rham cohomology classes of $(p|q)$ pseudoforms.

Following the discussion in \cite{Castellani:2015paa,Castellani:2017ycm}, we
need also the \textit{Picture Changing Operators} ${\mathbb{Y}}_{k}^{(0|1)}$.
They act multiplicatively (using the graded wedge product of pseudoforms) on
the spaces $\Omega^{(p|q)}:$
\begin{equation}
{\mathbb{Y}}_{k}^{(0|1)}:\Omega^{(p|q)}\longrightarrow\Omega^{(p|q+1)}\,,
\end{equation}
with $\omega^{(p|q+1)}=\omega^{(p|q)}\wedge{\mathbb{Y}}_{k}^{(0|1)}$. There
are here four possible independent directions along which ${\mathbb{Y}}%
_{k}^{(0|1)}$ can act, labelled by the index $k$. This means, for
example, that ${\mathbb{Y}}_{\alpha}^{(0|1)}$ is proportional to $\delta
(\psi^{\alpha})$, and ${\mathbb{Y}}_{\dot{\alpha}}^{(0|1)}$ is proportional to
$\delta(\bar{\psi}^{\dot{\alpha}})$. We denote by $\mathbb{Y}^{(0|4)}$
the product of four PCO's along the four possibile independent directions. As
discussed for example in \cite{Witten:2012bg} and \cite{Catenacci:2010cs}, the
product of two delta's is anticommuting (e.g. for $\delta(\psi^{1})\wedge
\delta(\bar{\psi}^{\dot{2}})=-\delta(\bar{\psi}^{\dot{2}})\wedge\delta
(\psi^{1}))$, guaranteeing that  no singularity arises when multiplying two or
more PCO's. Thus ${\mathbb{Y}}_{1}^{(0|1)}\wedge{\mathbb{Y}%
}_{1}^{(0|1)}=0$, etc.

As discussed in \cite{Castellani:2015paa,Castellani:2017ycm}, the PCO's of
type ${\mathbb{Y}}$ represent the Poincar\'{e} form dual to the embedding of
the reduced bosonic submanifold $\mathcal{M}^{(4|0)}$ into the supermanifold
$\mathcal{M}^{(4|4)}$. They are elements of the de Rham cohomology with the
properties
\begin{equation}
d{\mathbb{Y}}_{k}^{(0|1)}=0\,,~~~~~~{\mathbb{Y}}_{k}^{(0|1)}\neq d{\eta}%
_{k}^{(-1|1)}\,,~~~~\delta{\mathbb{Y}}_{k}^{(0|1)}=d{\Lambda}_{k}^{(-1|1)}%
\end{equation}
The last equation means that any variation (under a diff.) of the PCO is $d$-exact. This gives
\begin{equation}
d\omega^{(p|q+1)}=d\left[  \omega^{(p|q)}\wedge{\mathbb{Y}}_{k}^{(0|1)}%
\right]  =d\omega^{(p|q)}\wedge{\mathbb{Y}}_{k}^{(0|1)}%
\end{equation}
which implies also that ${\mathbb{Y}}_{k}^{(0|1)}$ maps cohomology classes into
cohomology classes:%
\begin{equation}
{\mathbb{Y}}_{k}^{(0|1)}:H_{d}(\Omega^{(p|q)})\longrightarrow H_{d}%
(\Omega^{(p|q+1)})
\end{equation}
The explicit form of ${\mathbb{Y}}_{k}^{(0|1)}$ is important in the
applications and we will elaborate on it in the forthcoming sections. In
particular there are choices with manifest symmetries, playing a crucial
r\^{o}le in building manifestly supersymmetric actions.

To decrease the picture, we use a different PCO operator denoted by $\mathbb{Z}%
_{k}^{(0|-1)}$ , acting as a double differential operator on the space of
pseudoforms
\begin{equation}
\mathbb{Z}_{k}^{(0|-1)}:\Omega^{(p|q)}\longrightarrow\Omega^{(p|q-1)}\,
\end{equation}
These operators act along different directions $k$
by removing the corresponding delta forms of type $\delta(\psi^{\alpha})$ or
$\delta(\bar{\psi}^{\dot{\alpha}})$. A convenient way to represent
$\mathbb{Z}_{k}^{(0|-1)}$ is given by
\begin{equation}
\mathbb{Z}_{k}^{(0|-1)}=\left[  d,\Theta(\iota_{k})\right]  =\delta(\iota
_{k}) \ell_{k} \label{dcN}%
\end{equation}
(see for examples again \cite{Castellani:2017ycm}) where $\Theta(\iota_{k})$
is the Heaviside step function and $\iota_{k}$ is the contraction along the
$\psi^{\alpha}$ or $\bar{\psi}^{\dot{\alpha}}$. Notice that $\Theta(\iota
_{k})$ is not a compact-support distribution and therefore it has to be
treated carefully. Nonetheless the explicit form of (\ref{dcN}) shows that
$\mathbb{Z}_{k}^{(0|-1)}$ is expressed only in terms of compact-support
distributions. $\ell_{k}$ is the Lie derivative along one of the vector
fields $D_{\alpha}$ or $\bar{D}_{\dot{\alpha}}$. The form (\ref{dcN}) is
computationally convenient when it acts on closed forms as will be seen
later. In addition, we also notice that the formula (\ref{dcN}) shows that the
operator $\mathbb{Z}_{k}^{(0|-1)}$ is ``closed" but it fails to be ``exact"
since $\Theta(\iota_{k})$ is not a compact-support distribution.

\subsection{Lorentz transformations on $\Omega^{(p|q)}$}

Before discussing in detail some of the relevant spaces $\Omega^{(p|q)}$ , we
would like to clarify how the Lorentz symmetry is implemented in the complex
of pseudoforms. This is a crucial point in order to understand how the
covariance is recovered at any picture number.

Let us consider an infinitesimal Lorentz transformation $\Lambda_{~b}^{a}$ of
$SO(3,1)$. It acts on the coordinates $x^{a},\theta^{\alpha}$,$\bar{\theta
}^{\dot{\alpha}}$ linearly according to vector and spinor representations
\begin{equation}
\delta x^{a}=\Lambda_{~b}^{a}x^{b}\,,~~~~~~\delta\theta^{\alpha}=\frac{1}%
{4}\Lambda_{ab}(\gamma^{ab})_{~\beta}^{\alpha}\theta^{\beta}\,,~~~~~~\delta
\bar{\theta}^{\dot{\alpha}}=\frac{1}{4}\Lambda_{ab}(\gamma^{ab})_{~\dot{\beta
}}^{\dot{\alpha}}\bar{\theta}^{\dot{\beta}}\,. \label{LOCA}%
\end{equation}
In the same way, the $(1|0)$-superforms $(V^{a},\psi^{\alpha},\bar{\psi}%
^{\dot{\alpha}})$ transform, respectively, in the vector and in the spinor
representations. Thus, all forms belonging to the complex with zero picture,
namely $\Omega^{(p|0)}$, transform in the tensorial representations of each
single monomial. For example, given $\omega_{\lbrack ab](\alpha_{1}\dots
\alpha_{n})}V^{a}V^{b}\psi^{\alpha_{1}}\dots\psi^{\alpha_{n}}$, the components
$\omega_{\lbrack ab](\alpha_{1}\dots\alpha_{n})}(x,\theta)$ transform in
the anti-symmetrized product of the vector representation tensored
with $n$-symmetrized  spinor representation.

If we consider the complex of integral forms $\Omega^{(p|4)}$, and we perform
an infinitesimal Lorentz transformation, we have to use distributional
relations as for example
\begin{equation}
\delta(a\psi^{1}+b\psi^{2})\delta(c\psi^{1}+d\psi^{2})={\det\left(
\begin{array}
[c]{cc}%
a & b\\
c & d
\end{array}
\right)  ^{-1}}\delta(\psi^{1})\delta(\psi^{2})\,, \label{LOCB}%
\end{equation}
implying that the product of $\delta(\psi^{1})\delta(\psi^{2})$ transforms
as the inverse of a density. Therefore each monomial of the complex
$\Omega^{(p|4)}$ transforms according to a tensorial representation of the
Lorentz group. For example, a finite variation of an integral top form
$\omega^{(4|4)}=f(x,\theta)V^{4}\delta^{2}(\psi)\delta^{2}(\bar{\psi})$ gives
\begin{equation}
\omega^{(4|4)}\longrightarrow\frac{\det(\Lambda_{~b}^{a})}{\det(\Lambda
_{~\beta}^{\alpha})\det(\bar{\Lambda}_{~\dot{\beta}}^{\dot{\alpha}})}f\left(
\Lambda_{b}^{a}x^{a},\Lambda_{~\beta}^{\alpha}\theta^{\beta},\bar{\Lambda
}_{\dot{\beta}}^{\dot{\alpha}}\bar{\theta}^{\dot{\beta}}\right)  V^{4}%
\delta^{2}(\psi)\delta^{2}(\bar{\psi})
\end{equation}
where $\Lambda_{~\beta}^{\alpha}=\frac{1}{4}\Lambda_{ab}(\gamma^{ab})_{~\beta
}^{\alpha}$ and $\bar{\Lambda}_{~\dot{\beta}}^{\dot{\alpha}}=\frac{1}%
{4}\Lambda_{ab}(\gamma^{ab})_{~\dot{\beta}}^{\dot{\alpha}}$. Since $\Lambda$
is a Lorentz transformation, i.e.  $\Lambda \in SO(3,1)$, all determinants appearing
in the front factor are equal to one and the top form is invariant if
\begin{equation}
f\left(  \Lambda_{b}^{a}x^{a},\Lambda_{~\beta}^{\alpha}\theta^{\beta}%
,\bar{\Lambda}_{\dot{\beta}}^{\dot{\alpha}}\bar{\theta}^{\dot{\beta}}\right)
=f(x,\theta,\bar{\theta})\,
\end{equation}

Let us now consider the complexes of pseudoforms, for example at picture one:
$\Omega^{(p|1)}$ for any $p\in\mathbb{Z}$. As seen above, it is unbounded from
above and from below and each space is infinite dimensional. For a single
Dirac delta function $\delta(\psi^{1})$, we cannot use the distributional
identity (\ref{LOCB}), but we observe that
\begin{align}
\delta(\psi^{1})\longrightarrow &  \delta\Big(\psi^{1}+\frac{1}{4}\Lambda
_{ab}(\gamma^{ab})_{~\beta}^{1}\psi^{\beta}\Big)\label{LOCC}\\
&  =\left(  1-\frac{1}{4}\Lambda_{ab}(\gamma^{ab})_{~1}^{1}\right)
\delta(\psi^{1})+\frac{1}{4}\Lambda_{ab}(\gamma^{ab})_{~2}^{1}\psi^{2}%
\delta^{(1)}(\psi^{1})+\mathcal{O}(\Lambda^2)\nonumber
\end{align}
where $\delta^{(1)}(\psi_{1})$ is the first derivative of $\delta(\psi^{1})$
and we have neglected the infinitesimal terms. The first term is obtained by
using the rule $\psi^{1}\delta^{(1)}(\psi^{1})=-\delta(\psi^{1})$ and the
second term comes from the Taylor expansion of the delta function. Then, in order to implement the Lorentz symmetry in the space of
pseudoforms $\Omega^{(p|1)}$,  all the components in the expansion of a generic
superform in $\Omega^{(p|1)}$ are needed, and span an infinite dimensional space.

\subsection{Superfields, Volume Forms and Chiral Volume Forms}

A superfield $\Phi$ is a $(0|0)$-superform and it has the conventional
superfield properties. Its supersymmetry transformations are deduced from its
differential
\begin{equation}
\delta\Phi=\ell_{\epsilon}\Phi=\iota_{\epsilon}d\Phi\,. \label{SSA}%
\end{equation}
where $\epsilon$ is the constant supersymmetry parameter.

An important ingredient for the subsequent sections are the volume forms,
necessary to build integral forms and therefore integrable quantities on
the entire supermanifold without referring to a specific 
coordinate system. As in general relativity, where the use of
differential forms is a powerful tool to construct diff. invariant
objects, here the construction of integral forms is needed to have
superdiff. invariant objects, that are in
turn also invariant under rigid supersymmetry. For this reason, we provide
here some remarks concerning the real and the chiral volume forms.

The top integral forms of $\Omega^{(4|4)}$ are represented by
\begin{equation}
\omega^{(4|4)}=\Phi(x,\theta,\bar{\theta})\epsilon_{abcd}V^{a}\wedge
V^{b}\wedge V^{c}\wedge V^{d}\epsilon_{\alpha\beta}\delta(\psi^{a})\delta
(\psi^{\beta})\epsilon_{\dot{\alpha}\dot{\beta}}\delta(\psi^{\dot{\alpha}%
})\delta(\psi^{\dot{\beta}}) \label{caffaC}%
\end{equation}
Rewriting the supervielbein 
$E^{A}=(V^{a},\psi^{\alpha},\bar{\psi}^{\dot{\alpha}})$ on a curved basis:
\begin{align}
V^{a}  &  =E_{m}^{a}dx^{m}+E_{\mu}^{a}d\theta^{\mu}+E_{\dot{\mu}}^{a}%
d\bar{\theta}^{\dot{\mu}}\nonumber\\
\psi^{a}  &  =E_{m}^{a}dx^{m}+E_{\mu}^{a}d\theta^{\mu}+E_{\dot{\mu}}^{a}%
d\bar{\theta}^{\dot{\mu}}\nonumber\\
\bar{\psi}^{\dot{\alpha}}  &  =E_{m}^{\dot{\alpha}}dx^{m}+E_{\mu}^{\dot
{\alpha}}d\theta^{\mu}+E_{\dot{\mu}}^{\dot{\alpha}}d\bar{\theta}^{\dot{\mu}}
\label{SupA}%
\end{align}
we find also:
\begin{align}
\omega^{(4|4)}  &  =\Phi(x,\theta,\bar{\theta})V^{4}\delta^{4}(\psi
)\equiv\epsilon_{abcd}V^{a}\wedge V^{b}\wedge V^{c}\wedge V^{d}\epsilon
_{\alpha\beta}\delta(\psi^{a})\delta(\psi)\epsilon_{\dot{\alpha}\dot{\beta}%
}\delta(\psi^{\dot{\alpha}})\delta(\psi^{\dot{\beta}})\nonumber\\
&  =\Phi(x,\theta,\bar{\theta})\mathrm{Sdet}(E)d^{4}x\delta^{4}(d\theta)
\label{SupB}%
\end{align}
This $(4|4)$ form is trivially closed (being a top integral form),  and not exact if $\Phi(x,\theta,\bar{\theta
})\mathrm{Sdet}(E)\neq\mathrm{constant}$. Its supersymmetry variation is
\begin{equation}
\delta\omega^{(4|4)}=\ell_{\epsilon}\omega^{(4|4)}=d\left(
\iota_{\epsilon}\omega^{(4|4)}\right)
\end{equation}

Notice that if $\Phi\mathrm{Sdet(E)}=1$, the top form
$\omega^{(4|4)}$ cannot be regarded as the true volume form. Indeed
\begin{equation}
{\tilde \omega}^{(4|4)} \equiv V^{4}\delta^{4}(\psi)=\epsilon_{abcd}V^{a}\wedge V^{b}\wedge
V^{c}\wedge V^{d}\epsilon_{\alpha\beta}\delta(\psi^{\alpha})\delta(\psi
^{\beta})\epsilon_{\dot{\alpha}\dot{\beta}}\delta(\bar{\psi}^{\dot{\alpha}%
})\delta(\bar{\psi}^{\dot{\beta}})\,.
\end{equation}
is closed, but it is also exact as can be shown using the
relation
\begin{equation}
\epsilon_{\dot{\alpha}\dot{\beta}}\delta(\bar{\psi}^{\dot{\alpha}})\delta
(\bar{\psi}^{\dot{\beta}})=d\Big[\bar{\theta}^{\dot{\alpha}}\bar{\iota}%
_{\dot{\alpha}}\delta^{2}(\bar{\psi})\Big]\label{caffaB}%
\end{equation}
to write ${\tilde \omega}^{(4|4)}$ as
\begin{equation}
{\tilde \omega}^{(4|4)}=d\Big[V^{4}\delta^{2}(\psi)\wedge
\bar{\theta}^{\dot{\alpha}}\bar{\iota}_{\dot{\alpha}}\delta^{2}(\bar{\psi
})\Big]\,.
\end{equation}
so that
\begin{equation}
\int_{\mathcal{M}^{(4|4)}}{\tilde \omega}^{(4|4)}=0
\end{equation}
by Stokes theorem\footnote{For notations and the integration theory of
superfields and integral forms we refer mainly to
\cite{Witten:2012bg,Castellani:2017ycm}.
\par
Stokes theorem for integral forms integration is discussed in 
reference \cite{Witten:2012bg}.}. Nevertheless the form ${\tilde \omega}^{(4|4)}$ can
be used to construct integral forms that can be integrated on the entire
supermanifold. Given a superfield $\Phi(x,\theta,\bar{\theta})$ we have:
\begin{equation}
\int_{\mathcal{M}^{(4|4)}}\Phi(x,\theta,\bar{\theta}){\tilde \omega}^{(4|4)}%
=\int
_{\mathcal{M}^{(4)}}d^{4}x\left.  D^{2}\bar{D}^{2}\Phi\right\vert
_{\theta=\bar{\theta}=0}%
\end{equation}
which in general does not vanish if $\Phi$ is not constant.

We can construct the chiral volume forms as follows. Given $\omega
^{(4|2)}=V^{4}\delta^{2}(\psi)$ , $\overline{\omega}^{(4|\bar{2})}=V^{4}%
\delta^{2}(\bar{\psi})$ and the PCO's ${\mathbb{Y}}^{(0|2)}=\theta^{2}%
\delta^{2}(\psi)$ and $\overline{\mathbb{Y}}^{(0|2)}=\bar{\theta}^{2}%
\delta^{2}(\bar{\psi})$, we have
\begin{equation}
\omega_{C}^{(4|4)}=V^{4}\delta^{2}(\psi)\wedge\overline{\mathbb{Y}}%
^{(0|2)}\,,~~~~~~\overline{\omega}_{C}^{(4|4)}={\mathbb{Y}}^{(0|2)}\wedge
V^{4}\delta^{2}(\bar{\psi})\,,
\end{equation}
They are conjugated to each other. They are closed, and in fact are
exact. This can be easily seen by using again the equation (\ref{caffaB}). The
differential of $V^{a}$ produces one $\psi^{\alpha}$ -- annihilated by the
contraction $\iota_{\alpha}$ -- and one $\bar{\psi}^{\dot{\alpha}}$ which however
is not cancelled by the Dirac deltas $\delta^{2}(\bar{\psi})$ which are present
in $\omega^{(4|4)}$, but not in $\omega^{(4|2)}$. Then, we have
\begin{align}
&  \int_{\mathcal{M}^{(4|4)}}\Phi(x,\theta,\bar{\theta})\left(  V^{4}%
\delta^{2}(\psi)\wedge\overline{\mathbb{Y}}^{(0|2)}+{\mathbb{Y}}^{(0|2)}\wedge
V^{4}\delta^{2}(\bar{\psi})\right)  \left[  d^{4}xd^{2}\theta d^{2}\bar
{\theta}d^{2}\psi d^{2}\bar{\psi}\right]  =\nonumber\\
&  \int_{\mathcal{M}^{(4|2,0)}}\Phi(x,\theta,0)V^{4}\delta^{2}(\psi)\left[
d^{4}xd^{2}\theta d^{2}\psi\right]  +\int_{\mathcal{M}^{(4|0,2)}}%
\Phi(x,0,\bar{\theta})V^{4}\delta^{2}(\bar{\psi})\left[  d^{4}xd^{2}%
\bar{\theta}d^{2}\bar{\psi}\right]  =\nonumber\\
&  =\int_{\mathcal{M}^{(4)}}d^{4}x\left.  D^{2}\Phi\right\vert _{\theta
=0}+\int_{\mathcal{M}^{(4)}}d^{4}x\left.  \bar{D}^{2}\Phi\right\vert
_{\bar{\theta}=0}%
\end{align}
The result is a sum of a chiral and an anti-chiral term integrated over the
reduced bosonic submanifold of the supermanifold.


\subsection{Chevalley-Eilenberg Cohomology}

The next step is to analyze some other interesting sectors of the cohomology.
In particular those which are relevant for Wess-Zumino and super-Yang-Mills
actions. It turns out that the crucial ingredients for the forthcoming
sections are elements of the cohomology $H_{d}(\Omega^{(4|0)})$ with two
vectorial vielbeins and two spinorial vielbeins, i.e. with the
generic form:
\begin{equation}
\omega^{(4|0)} \sim\bar{\theta}\theta\,\bar{\psi}\wedge\psi\wedge V\wedge
V+\theta^{2}\bar{\psi}^{2}\wedge V\wedge V+\mathrm{h.c.}%
\end{equation}
These differential forms are dual to the PCO's listed in (\ref{NEWA}), in the sense 
that:
\begin{equation}
\omega^{(4|0)}\wedge{\mathbb{Y}}^{(0|4)}\sim\bar{\theta}^{2}\theta^{2}%
V^{4}\delta^{4}(\psi)\,.
\end{equation}
The factor $\bar{\theta}^2 \theta^{2}$ appearing in the r.h.s.  is
crucial in order to have a closed, but not exact, integral form.

Now, in order to find the appropriate expression for $\omega^{(4|0)}$ we list the
possible Lorentz invariant forms with two $\theta$'s and two $\psi$'s:
\begin{align}
\omega_{1}  &  =\frac{1}{2}\left(  \theta_{\alpha}\bar{\theta}_{\dot{\alpha}%
}\psi_{\beta}\bar{\psi}_{\dot{\beta}}-\theta_{\beta}\bar{\theta}_{\dot{\beta}%
}\psi_{\alpha}\bar{\psi}_{\dot{\alpha}}\right)  V^{\alpha\dot{\alpha}}\wedge
V^{\beta\dot{\beta}}=(\theta V\bar{\theta})(\psi V\bar{\psi})\,,\label{choC}\\
\omega_{2}  &  =\frac{1}{2}\left(  \theta_{\alpha}\bar{\theta}_{\dot{\beta}%
}\psi_{\beta}\bar{\psi}_{\dot{\alpha}}-\theta_{\beta}\bar{\theta}_{\dot
{\alpha}}\psi_{\alpha}\bar{\psi}_{\dot{\beta}}\right)  V^{\alpha\dot{\alpha}%
}\wedge V^{\beta\dot{\beta}}=(\theta V\bar{\psi})(\psi V\bar{\theta
})\,,\nonumber\\
\omega_{3}  &  =\frac{1}{2}(\theta_{\gamma}\psi^{\gamma})\left(  \bar{\theta
}_{\dot{\alpha}}\bar{\psi}_{\dot{\beta}}+\bar{\theta}_{\dot{\beta}}\bar{\psi
}_{\dot{\alpha}}\right)  \epsilon_{\alpha\beta}V^{\alpha\dot{\alpha}}\wedge
V^{\beta\dot{\beta}}=(\theta\cdot\psi)(\bar{\theta}V_{-}^{2}\bar{\psi
})\,,\nonumber\\
\omega_{4}  &  =\frac{1}{2}(\bar{\theta}_{\dot{\gamma}}\bar{\psi}^{\dot
{\gamma}})\left(  \theta_{\alpha}\psi_{\beta}+\theta_{\beta}\psi_{\alpha
}\right)  \epsilon_{\dot{\alpha}\dot{\beta}}\,V^{\alpha\dot{\alpha}}\wedge
V^{\beta\dot{\beta}}=(\bar{\theta}\cdot\bar{\psi})(\theta V_{+}^{2}%
\psi)\,,\nonumber\\
\omega_{5}  &  =\theta^{\gamma}\epsilon_{\gamma\rho}\theta^{\rho}(\bar{\psi
}_{\dot{\alpha}}\bar{\psi}_{\dot{\beta}})\epsilon_{\alpha\beta}V^{\alpha
\dot{\alpha}}\wedge V^{\beta\dot{\beta}}=\theta^{2}(\bar{\psi}V_{-}^{2}%
\bar{\psi})\nonumber\\
\omega_{6}  &  =\bar{\theta}^{\dot{\gamma}}\epsilon_{\dot{\gamma}\dot{\rho}%
}\bar{\theta}^{\dot{\rho}}(\psi_{\alpha}\psi_{\beta})\epsilon_{\dot{\alpha
}\dot{\beta}}V^{\alpha\dot{\alpha}}\wedge V^{\beta\dot{\beta}}=\bar{\theta
}^{2}(\psi V_{+}^{2}\psi)\,,\nonumber
\end{align}
We have defined:
\begin{align}
&  (\theta V\bar{\theta})=\theta^{\alpha}V^{\beta{\dot{\beta}}}\bar{\theta
}^{{\dot{\alpha}}}\epsilon_{\alpha\beta}\epsilon_{{\dot{\beta}}{\dot{\alpha}}%
}\,,~~~~~~~~(\psi V\bar{\psi})=\psi^{\alpha}V^{\beta{\dot{\beta}}}\bar{\psi
}^{{\dot{\alpha}}}\epsilon_{\alpha\beta}\epsilon_{{\dot{\beta}}{\dot{\alpha}}%
}\,,\label{choCA}\\
&  (\theta V\bar{\psi})=\theta^{\alpha}V^{\beta{\dot{\beta}}}\bar{\psi}%
^{{\dot{\alpha}}}\epsilon_{\alpha\beta}\epsilon_{{\dot{\beta}}{\dot{\alpha}}%
}\,,~~~~~~~~(\psi V\bar{\theta})=\psi^{\alpha}V^{\beta{\dot{\beta}}}%
\bar{\theta}^{{\dot{\alpha}}}\epsilon_{\alpha\beta}\epsilon_{{\dot{\beta}%
}{\dot{\alpha}}}\,,\nonumber\\
&  (\theta\cdot\psi)=\theta^{\alpha}\psi^{\beta}\epsilon_{\alpha\beta
}\,,~~~~~~~~~~~~~~~(\bar{\theta}V_{-}^{2}\bar{\psi})=\bar{\theta}%
^{{\dot{\alpha}}}(V_{-}^{2})^{{\dot{\beta}}{\dot{\gamma}}}\bar{\psi}%
^{\dot{\delta}}\epsilon_{{\dot{\alpha}}{\dot{\beta}}}\epsilon_{{\dot{\gamma}%
}\dot{\delta}}\,,\nonumber\\
&  (\bar{\theta}\cdot\bar{\psi})=\bar{\theta}^{{\dot{\alpha}}}\bar{\psi
}^{{\dot{\beta}}}\epsilon_{{\dot{\alpha}}{\dot{\beta}}}%
\,,~~~~~~~~~~~~~~~(\theta V_{+}^{2}\psi)=\theta^{\alpha}(V_{+}^{2}%
)^{\beta\gamma}\psi^{\delta}\epsilon_{\alpha\beta}\epsilon_{\gamma\delta
}\,,\nonumber\\
&  \theta^{2}=\theta^{\alpha}\theta^{\beta}\epsilon_{\alpha\beta
}\,,~~~~~~~~~~~~~~~~~~~~~~~~\bar{\theta}^{2}=\bar{\theta}^{{\dot{\alpha}}}%
\bar{\theta}^{{\dot{\beta}}}\epsilon_{{\dot{\alpha}}{\dot{\beta}}%
}\,,\nonumber\\
&  (\bar{\psi}V_{-}^{2}\bar{\psi})=\bar{\psi}^{{\dot{\alpha}}}(V_{-}%
^{2})^{{\dot{\beta}}{\dot{\gamma}}}\bar{\psi}^{\dot{\delta}}\epsilon
_{{\dot{\alpha}}{\dot{\beta}}}\epsilon_{{\dot{\gamma}}\dot{\delta}}\,,~~(\psi
V_{+}^{2}\psi)=\psi^{\alpha}(V_{+}^{2})^{\beta\gamma}\psi^{\delta}%
\epsilon_{\alpha\beta}\epsilon_{\gamma\delta}\nonumber
\end{align}
whose differentials are
\begin{align}
d(\theta V\bar{\theta})  &  =(\psi V\bar{\theta})+(\theta V\bar{\psi
})-2i(\theta\cdot\psi)(\bar{\psi}\cdot\bar{\theta})\label{choCB}\\
d(\theta V\bar{\psi})  &  =(\psi V\bar{\psi})\,,~~~~~~~d(\psi V\bar{\theta
})=-(\psi V\bar{\psi})\nonumber\\
d(\psi V\bar{\psi})  &  =0\,,~~~~~~~~d(\psi V_{+}^{2}\psi)=0\,,\nonumber\\
d(\bar{\psi}V_{-}^{2}\bar{\psi})  &  =0\,,~~~~~~d(\theta\cdot\psi
)=0\,,~~~~~~~d(\bar{\theta}\cdot\bar{\psi})=0\,,\nonumber
\end{align}
The linear combination
\begin{equation}
\omega^{(4|0)} =a\,\omega_{1}+b\,\omega_{2}+c\,\omega_{3}+d\,\omega_{4}+e\,\omega
_{5}+f\,\omega_{6}%
\end{equation}
is closed if $a=c-d,e=f$ and $b=\frac{1}{2}(c+d)+2e$. If, in addition, we
require the hermiticity of $\omega$, one finds $c=d$ and therefore $a=0$. Then, we
find that the combination
\begin{equation}
\omega^{(4|0)} =c(\omega_{2}+\omega_{3}+\omega_{4})+e(2\omega_{2}+\omega_{5}+
\omega_{6})\,,
\end{equation}
is closed, real, and depends upon the two parameters $c$ and $e$.
Furthermore, we have to check whether this expression is exact. We observe
that there is only one  real candidate
(with $r$ a real parameter):
\begin{equation}
\gamma^{(3|0)}=r\Big(\theta^{2}(\bar{\psi}V_{-}^{2}\bar{\theta})+\bar{\theta
}^{2}(\psi V_{+}^{2}\theta)\Big)\,.
\end{equation}
such that $d\gamma^{(3|0)}$ has a structure similar to the ones listed in 
in (\ref{choC}).
Computing $d\gamma^{(3|0)}$  and adding it to $\omega^{(4|0)}$, we
finally end up with the expression
\begin{equation}
\omega^{(4|0)}=(c+2e)\omega_{2}+(c-2r)(\omega_{3}+\omega_{4})+(e+r)(\omega
_{5}+\omega_{6})
\end{equation}
and we can use the parameter $r$ to set one of the two combinations to zero.
If we choose $c=2r$, we see that the full expression is proportional to
$(c+2e)$. In the same way by choosing $r=-e$, we obtain again an expression which is
proportional to the combination $(c+2e)$. Therefore, after subtracting
the exact piece, we get a single representative in the cohomology class.

Notice that $\omega^{(4|0)}$ is not manifestly supersymmetric since it depends
upon $\theta$ and $\bar\theta$. This is the reason why this cohomology was
never used. However, its supersymmetry variation is $d$-exact.

\subsection{The PCO's $\mathbb{Y}^{(0|1)}$}

The easiest example of PCO that we can build is the one that projects the
theory on the bosonic submanifold by switching off the $\theta$ coordinates
and their differentials. For each coordinate we have the following four PCO's
acting along the $\theta$-directions
\begin{equation}
\mathbb{Y}_{1}^{(0|1)}=\theta^{1}\delta(\psi^{1})\,,~~~~\mathbb{Y}_{2}%
^{(0|1)}=\theta^{2}\delta(\psi^{2})\,,~~~~\mathbb{Y}_{\dot{1}}^{(0|1)}%
={\bar{\theta}}^{\dot{1}}\delta(\bar{\psi}^{\dot{1}})\,,~~~~\mathbb{Y}%
_{\dot{2}}^{(0|1)}={\bar{\theta}}^{\dot{2}}\delta(\bar{\psi}^{\dot{2}})\,.
\label{ppoA}%
\end{equation}
Each of them increases by one the picture of the form and projects to zero
the corresponding coordinate. Notice that they have a non-trivial kernel, for
example the kernel $\mathbb{Y}_{1}^{(0|1)}$ consists of linear functions of
$\theta^{1}$, $\psi^{1}$ and $\delta(\psi^{1})$ (due to the anticommutation
properties of the deltas). All PCO's in (\ref{ppoA}) are closed and not exact.
They are invariant under partial supersymmetry (for example $\mathbb{Y}%
_{1}^{(0|1)}$ is invariant under the supersymmetries along $\theta^{2}%
,{\bar{\theta}}^{\dot{1}}$ and ${\bar{\theta}}^{\dot{2}}$). As already
noticed, its supersymmetry variation is exact. The wedge product of all four
PCO's produces a single operator (up to an overall sign) which we denote by
\begin{equation}
\mathbb{Y}^{(0|4)}=\theta^{2}\delta^{2}(\psi)\bar{\theta}^{2}\delta^{2}%
(\bar{\psi})
\end{equation}
This PCO is trivially closed, it is not exact and it is not
manifestly supersymmetric. Nonetheless, its supersymmetry transformation is
$d$-exact. Therefore, given a closed superform $\mathcal{L}^{(4|0)}$, we can
write an action
\begin{equation}
S=\int_{\mathcal{M}^{(4|4)}}\mathcal{L}^{(4|0)}\wedge\mathbb{Y}^{(0|4)}
\label{PCO_B}%
\end{equation}
which reduces to the component action (which means the integral of
$\mathcal{L}^{(4|0)}$ computed at $\theta=\bar{\theta}=0$ and $\psi=\bar{\psi
}=0$ over $\mathcal{M}^{(4)}$).

The closure of $\mathcal{L}^{(4|0)}$ guarantees the supersymmetry invariance
of the action up to boundary terms. A milder condition can be imposed on
$\mathcal{L}^{(4|0)}$ in order for $S$ to be supersymmetric invariant:
\begin{equation}
\iota_{\epsilon}d\mathcal{L}^{(4|0)}=d \xi
\end{equation}
i.e. the differential along the supersymmetry directions
must be exact. The computation of the integral in (\ref{PCO_B}) along the
$\theta$'s and the $\psi$'s leads to
\begin{equation}
S=\int_{\mathcal{M}^{(4)}}\left.  \mathcal{L}^{(4|0)}\right\vert
_{\theta=0,\psi=0}\,,
\end{equation}
which is the component action and it is supersymmetric invariant if the
supersymmetry variation of the Lagrangian $\left.  \mathcal{L}^{(4|0)}%
\right\vert _{\theta=0,\psi=0}$ is an exact differential.

To rewrite the action in a manifestly supersymmetric way, we need another PCO
which is manifestly supersymmetric. It should have picture number equal to $4$
and zero form degree. To get from the Lagrangian $\mathcal{L}^{(4|0)}$ a top
integral form, the PCO should be closed, not exact, and possibly invariant
under supersymmetry. For that purpose, we consider the following six
combinations
\begin{align}
Y_{1}  &  =V^{\alpha\dot{\alpha}}\wedge V^{\beta\dot{\beta}}\left(
\theta_{\alpha}\bar{\theta}_{\dot{\alpha}}\iota_{\beta}\bar{\iota}_{\dot
{\beta}}-\theta_{\beta}\bar{\theta}_{\dot{\beta}}\iota_{\alpha}\bar{\iota
}_{\dot{\alpha}}\right)  \delta^{4}(\psi)\,,\nonumber\\
Y_{2}  &  =V^{\alpha\dot{\alpha}}\wedge V^{\beta\dot{\beta}}\left(
\theta_{\alpha}\bar{\theta}_{\dot{\beta}}\iota_{\beta}\bar{\iota}_{\dot
{\alpha}}-\theta_{\beta}\bar{\theta}_{\dot{\alpha}}\iota_{\alpha}\bar{\iota
}_{\dot{\beta}}\right)  \delta^{4}(\psi)\,,\nonumber\\
Y_{3}  &  =V^{\alpha\dot{\alpha}}\wedge V^{\beta\dot{\beta}}\epsilon
_{\alpha\beta}\left(  \bar{\theta}_{\dot{\alpha}}\bar{\iota}_{\dot{\beta}%
}+\bar{\theta}_{\dot{\beta}}\bar{\iota}_{\dot{\alpha}}\right)  \,\theta
^{\gamma}\iota_{\gamma}\delta^{4}(\psi)\,,\nonumber\\
Y_{4}  &  =V^{\alpha\dot{\alpha}}\wedge V^{\beta\dot{\beta}}\epsilon
_{\dot{\alpha}\dot{\beta}}\left(  \theta_{\alpha}\iota_{\beta}+\theta_{\beta
}\iota_{\alpha}\right)  \,\bar{\theta}^{\dot{\gamma}}\iota_{\dot{\gamma}%
}\delta^{4}(\psi)\,,\nonumber\\
Y_{5}  &  =V^{\alpha\dot{\alpha}}\wedge V^{\beta\dot{\beta}}\epsilon
_{\alpha\beta}(\bar{\theta}_{\dot{\gamma}}\bar{\theta}^{\dot{\gamma}}%
)\,\bar{\iota}_{\dot{\alpha}}\bar{\iota}_{\dot{\beta}}\delta^{4}%
(\psi)\,,\nonumber\\
Y_{6}  &  =V^{\alpha\dot{\alpha}}\wedge V^{\beta\dot{\beta}}\epsilon
_{\dot{\alpha}\dot{\beta}}(\theta_{\gamma}\theta^{\gamma})\,\iota_{\alpha
}\iota_{\beta}\delta^{4}(\psi)\,. \label{NEWA}%
\end{align}
The six possible forms reproduce the terms appearing in the Lagrangian 
(\ref{WZE}) (see the $\psi\bar{\psi}W\bar{W}VV$ terms in sec. 2).

They are indeed the
terms needed to reproduce the full superspace action. The two contraction
operators $\iota\bar{\iota}$ appearing in the operators act on the Lagrangian
by selecting the terms proportional to the combination $\bar{\psi}\psi$. In
addition, the factors $\theta\bar{\theta}$ are needed to prevent the PCO 
being exact.

By adjusting the six constants $a_{i}$ we can make the combination
\begin{equation}
{\mathbb{Y}}^{(0|4)}=\sum_{i=1}^{6}a_{i}Y_{i}\,,
\end{equation}
closed. Let us first impose the hermiticity by setting $a_{3}=a_{4}$ and
$a_{5}=a_{6}$. This reduces the structures to the four combinations
$Y_{1},Y_{2},Y_{3}+Y_{4},Y_{5}+Y_{6}$. Imposing the closure, we get $a_{1}=0$,
$a_{2}=-2(a_{3}+a_{5})$. Therefore, there are two independent structures which
are closed. However, there is a combination which is also exact. 

This can be easily derived by computing the variation of 
\begin{eqnarray}
\label{fanculo}
\eta^{(-1|4)} =\Big(
\theta^2 \bar\theta \cdot \bar \iota (\iota V^2_+ \iota) + \bar\theta^2 \theta\cdot \iota (\bar \iota V^2_- \bar\iota)\Big) \delta^4(\psi)\,.
\end{eqnarray}
Therefore, to
select a representative of the cohomology class we fix one of the coefficients,
avoiding the exact combination. For example we can set $a_{3}=0$ to simplify the structure as much as
possible:
\begin{equation}
{\mathbb{Y}}^{(0|4)}=\Big(-4(\theta V\bar{\iota})\wedge(\bar{\theta}%
V\iota)+\theta^{2}(\iota V\wedge V\iota)+\bar{\theta}^{2}(\bar{\iota}V\wedge
V\bar{\iota})\Big)\delta^{4}(\psi) \label{NEWBA}%
\end{equation}
Notice that there is a single non-chiral and two chiral and anti-chiral terms.
This already suggests how the three terms of the action in superspace emerge
from the geometrical action.

\subsection{The PCO's $\mathbb{Z}^{(0|-1)}$}

We have seen that  the complexes of pseudoforms are connected by the picture changing
operators. In the previous section we also observed that there are
some non-trivial cohomology classes needed for physics applications.
We check here that these cohomology classes are related by the PCO's.

Let us first analyze the action of $\mathbb{Z}^{(0|-1)}$ on the chiral forms.

They are supersymmetric invariant and we can apply the PCO's $Z_{\alpha
}=\left[  d,\Theta(\iota_{\alpha})\right]  $ to get the image in
$\Omega^{(4|0)}$. Since this computation is very instructive we report it here
in some detail. We have to act with the PCO's as follows:
\begin{align}
Z_{1}\Big(V^{4}\delta^{2}(\psi)\Big)  &  =\left[  d,\Theta(\iota_{1})\right]
V^{4}\delta^{2}(\psi)=d\Big[\Theta(\iota_{1})V^{4}\delta^{2}(\psi
)\Big]\label{SupO}\\
&  =d\left[  \frac{V^{4}}{\psi^{1}}\delta(\psi^{2})\right]  =\Big(\bar{\psi
}^{\dot{1}}V^{1\dot{2}}V^{2\dot{1}}V^{2\dot{2}}-\bar{\psi}^{\dot{2}}%
V^{1\dot{1}}V^{2\dot{1}}V^{2\dot{2}}\Big)\delta(\psi^{2})\,.
\end{align}
Notice that the result does not contain inverse powers\footnote{Negative
powers of the forms $\psi$ exist and are well defined only in picture $0$. In
this case the inverses of the $\psi^{\prime}s$ are closed and exact and behave
as negative degree superforms. The enlarged modules that contain also these
inverses extend to the left the complex of superforms (the first line in
figure 1). In picture $\neq0$ negative powers are not defined because of the
distributional relation $\psi\delta\left(  \psi\right)  =0.$} of $\psi$'s. In
the same way, we have $Z_{2}=[d,\Theta(\iota_{2})]$ and
\begin{align}
&  Z_{2}\Big(\bar{\psi}^{\dot{1}}V^{1\dot{2}}V^{2\dot{1}}V^{2\dot{2}}%
-\bar{\psi}^{\dot{2}}V^{1\dot{1}}V^{2\dot{1}}V^{2\dot{2}}\Big)\delta(\psi
^{2})\label{SupO1}\\
&  =d\left[  \frac{1}{\psi^{2}}\Big(\bar{\psi}^{\dot{1}}V^{1\dot{2}}%
V^{2\dot{1}}V^{2\dot{2}}-\bar{\psi}^{\dot{2}}V^{1\dot{1}}V^{2\dot{1}}%
V^{2\dot{2}}\Big)\right] \\
&  =\left[  \bar{\psi}^{\dot{1}}\bar{\psi}^{\dot{2}}(V^{1\dot{2}}V^{2\dot{1}%
}+V^{1\dot{1}}V^{2\dot{2}})-(\bar{\psi}^{\dot{1}})^{2}V^{1\dot{2}}V^{2\dot{2}%
}-(\bar{\psi}^{\dot{2}})^{2}V^{1\dot{1}}V^{2\dot{1}}\right] \\
&  =V^{\alpha\dot{\alpha}}\wedge V^{\beta\dot{\beta}}\epsilon_{\alpha\beta
}\bar{\psi}_{\dot{\alpha}}\bar{\psi}_{\dot{\beta}}\,.
\end{align}
(with $\bar{\psi}_{\dot{\alpha}}=\epsilon_{\dot{\alpha}\dot{\beta}}\bar{\psi
}^{\dot{\beta}}$). This form is closed, supersymmetric invariant and
polynomial in $V^{a},\psi^{a}$ and $\bar{\psi}^{\dot{\alpha}}$ (this means
that it is indeed a superform). Notice that we get only the chiral part of the
cohomology of $\Omega^{(4|0)}$. Starting from the antichiral integral form
$V^{4}\delta^{2}(\bar{\psi})$, we would get the other class in $H_{d}^{(4|0)}$.

We consider now the following volume form where we have chosen $\Phi
(x,\theta,\bar{\theta})$ in (\ref{caffaC}) to be equal to the product of the
$\theta$'s and $\bar{\theta}$'s,
\begin{eqnarray}
\label{volumazza}
\mathrm{Vol}^{(4|4)}=V^{4}\theta^{1}\theta^{2}\bar{\theta}^{\dot{1}}%
\bar{\theta}^{\dot{2}}\,\delta^{2}(\psi)\delta^{2}({\bar{\psi}})
\end{eqnarray}
and where we have written the spinorial indices explicitly to simplify the
derivation. We use the notations in {app. B} for the product of the vielbeins.

We act with the PCO $Z_{1}=[d,\Theta(\iota_{D_{1}})]$ on the volume form:
\begin{align}
Z_{1}\mathrm{Vol}^{(4|4)}  &  =[d,\Theta(\iota_{D_{1}})]\mathrm{Vol}%
^{(4|4)}=d\left[  \Theta(\iota_{D_{1}})\mathrm{Vol}^{(4|4)}\right]
\label{camB}\\
&  =d\left[  V^{4}\theta^{1}\theta^{2}\bar{\theta}^{\dot{1}}\bar{\theta}%
^{\dot{2}}\,\frac{1}{\psi_{1}}\delta(\psi_{2})\delta^{2}({\bar{\psi}})\right]
=V^{4}\theta^{2}\bar{\theta}^{\dot{1}}\bar{\theta}^{\dot{2}}\,\delta(\psi
_{2})\delta^{2}({\bar{\psi}})\,.
\end{align}
Acting with $Z_{2}=[d,\Theta(\iota_{D_{2}})]$, we find
\[
Z_{2}Z_{1}\mathrm{Vol}^{(4|4)}=[d,\Theta(\iota_{D_{2}})]V^{4}\theta^{2}%
\bar{\theta}^{\dot{1}}\bar{\theta}^{\dot{2}}\,\delta(\psi_{2})\delta^{2}%
({\bar{\psi}})=V^{4}\bar{\theta}^{\dot{1}}\bar{\theta}^{\dot{2}}\,\delta
^{2}({\bar{\psi}})
\]
This form is the  chiral volume form which is closed and not exact. To
proceed, we can act with the PCO removing the $\delta$'s depending on
$\bar{\psi}$'s:
\begin{align}
\bar{Z}_{1}Z_{2}Z_{1}\mathrm{Vol}^{(4|4)}  &  =d\left[  V^{4}\bar{\theta
}^{\dot{1}}\bar{\theta}^{\dot{2}}\frac{1}{\bar{\psi}_{\dot{1}}}\delta
(\bar{\psi}_{\dot{2}})\right] \label{camD}\\
&  =\left[  \psi_{\alpha}(V_{3})^{\alpha\dot{\alpha}}\bar{\psi}_{\dot{\alpha}%
}\,\bar{\theta}^{\dot{2}}\frac{1}{\bar{\psi}_{\dot{1}}}\delta(\bar{\psi}%
_{\dot{2}})+V^{4}\bar{\theta}\bar{\theta}^{\dot{2}}\delta(\bar{\psi}_{\dot{2}%
})\right] \nonumber\\
&  =\left[  \psi_{\alpha}(V_{3})^{\alpha\dot{1}}\bar{\theta}^{\dot{2}}%
\delta(\bar{\psi}_{\dot{2}})+V^{4}\bar{\theta}^{\dot{2}}\delta(\bar{\psi
}_{\dot{2}})\right] \nonumber
\end{align}
where all the inverse powers of $\bar{\psi}$'s disappeared. For the last step,
we act with $\bar{Z}_{2}$, and we have
\begin{align}
\bar{Z}_{2}\bar{Z}_{1}Z_{2}Z_{1}\mathrm{Vol}^{(4|4)}  &  =d\left[
\psi_{\alpha}(V_{3})^{\alpha\dot{1}}\bar{\theta}^{\dot{2}}\frac{1}{\bar{\psi
}_{\dot{2}}}+V^{4}\bar{\theta}^{\dot{2}}\frac{1}{\bar{\psi}_{\dot{2}}}\right]
\label{camE}\\
&  =-\frac{i}{2}(\psi V_{+}^{2}\psi)\bar{\theta}^{2}+\psi_{\alpha}%
(V_{3})^{\alpha\dot{\alpha}}\bar{\theta}_{\dot{\alpha}}+V_{4}\nonumber
\end{align}
where $\bar{\theta}_{\dot{\alpha}} = \epsilon_{\dot\alpha\dot\beta}
\bar{\theta}^{\dot{\beta}}$.

One obtains a covariant expression since all indices are suitably contracted. In
the same way, one could act first with the $\bar{Z}$'s and then with the $Z$'s
to find
\begin{equation}
Z_{2}Z_{1}\bar{Z}_{2}\bar{Z}_{1}\mathrm{Vol}^{(4|4)}=\frac{i}{2}(\psi
V_{-}^{2}\psi)\theta^{2}+\bar{\psi}_{\dot{\alpha}}(V_{3})^{\alpha\dot{\alpha}%
}\theta_{\alpha}+V_{4} \label{camF}%
\end{equation}
where $\theta_{\alpha} = \epsilon_{\alpha\beta} \theta^{\beta}$.

Note that we can relate the two formulae above by observing that:
\begin{equation}
d[\theta_{\alpha}(V_{3})^{\alpha\dot{\alpha}}\bar{\theta}_{\dot{\alpha}%
}]=i\theta_{\alpha}(V_{3})^{\alpha\dot{\alpha}}\bar{\theta}_{\dot{\alpha}%
}-i\psi_{\alpha}(V_{3})^{\alpha\dot{\alpha}}\bar{\psi}_{\dot{\alpha}}%
+i(\theta\cdot\psi)(\bar{\theta}V_{-}^{2}\bar{\psi})-i(\bar{\theta}\cdot
\bar{\psi})(\theta V_{+}^{2}\psi)
\end{equation}
which allows us to rewrite the second term in (\ref{camE}) as the second term
in (\ref{camF}). Combining the two expressions we end up with the final
result
\begin{equation}
\bar{Z}_{2}\bar{Z}_{1}Z_{2}Z_{1}\mathrm{Vol}^{(4|4)}+Z_{2}Z_{1}\bar{Z}_{2}%
\bar{Z}_{1}\mathrm{Vol}^{(4|4)}=\omega^{(4|0)}+d\eta
\end{equation}
where $\omega^{(4|0)}$ is in the Chevalley-Eilenberg cohomology class
discussed above. Thus, we have shown that acting with the PCO's $Z$ on the 
volume form (\ref{volumazza}) reproduces the Chevalley-Eilenberg cohomology 
discussed in the previous sections. Notice that the presence of $\theta$'s and $\bar \theta$'s 
is essential to reconstruct the cohomology by acting with PCO's. 

\subsection{Two useful theorems}

As an application of the previous discussions, we illustrate in this section
two theorems playing an important r\^{o}le in the superspace analysis of physical theories 
(see also \cite{Buchbinder:1998qv}).

The first is an application of Stokes theorem to supermanifolds with
torsion, and is very useful in manipulating the superspace
Lagrangians since it simplifies many computations. The use of integral forms
is very well adapted to such manipulations since Stokes' theorem is valid
for integral forms (and is a strong motivation for their
integration theory) and well-known techniques can be employed here.

The second theorem is very useful for treating supergravity theories. In that
framework some important quantities, such as the Ricci scalar or the Riemann
tensor, appear in the superspace expansion of some superfields. Therefore,
disentangling those physical components from a given superfield is crucial for
building  actions. One important example is the relation between
curved chiral and anti-chiral volume forms with the Ricci scalar of the
manifold and the non-chiral volume form. We show that this is very natural in
the context of integral forms where the volume form plays an essential r\^{o}le.

In studying the relation between the chiral volume forms and the non-chiral
one, we face the problem of computing the variation of the superdeterminant of
the supervielbein. For that purpose, we use the integral forms for a straight derivation.

We recall that, if we denote by $\nabla_{A}$ the supercovariant derivative
(w.r.t. the spin connection $\omega^{ab}$), we have the equations
\begin{align}
&  \nabla_{A}V^{a}=T_{~Ab}^{a}V^{b}+T_{~A\beta}^{a}\psi^{\beta}%
\,,~~~~~~\label{theoA}\\
&  \nabla_{A}\psi^{\alpha}=T_{~Ab}^{\alpha}V^{b}+T_{~A\beta}^{\alpha}%
\psi^{\beta}\,,~~~~~~\nonumber\\
&  \nabla_{A}\omega^{ab}+\omega_{A,c}^{a}\omega^{cb}=R_{~~Ac}^{ab}%
V^{c}+R_{~~A\beta}^{ab}\psi^{\beta}\,,\nonumber
\end{align}
where $T_{~~BC}^{A}$ are the components of the torsion $T^{a}=\frac{1}%
{2}T_{~~AB}^{a}E^{A}\wedge E^{B}$ and where $E^{A}=(V^{a},\psi^{\alpha})$ (we
do not impose any constraints and we use the greek indices to denote the $4$
spinors components in the Majorana representation).

We act with $\nabla_{A}$ on $\omega^{(4|4)}$ as follows
\begin{align}
\nabla_{A}\omega^{(4|4)} &  =\nabla_{A}\Big(\epsilon_{abcd}V^{a}\dots
V^{b}\delta^{4}(\psi)\Big)\label{theoB}\\
&  =4\epsilon_{abcd}(\nabla_{A}V^{a})\dots V^{d}\delta^{4}(\psi)+\epsilon
_{abcd}V^{a}\dots V^{d}(\nabla_{A}\psi^{\alpha})\iota_{\alpha}\delta^{4}%
(\psi)\nonumber\\
&  =4\epsilon_{abcd}\Big(T_{~Ae}^{a}V^{e}+T_{~A\beta}^{a}\psi^{\beta
}\Big)\dots V^{d}\delta^{4}(\psi)+\epsilon_{abcd}V^{a}\dots V^{d}%
\Big(T_{~Ae}^{\alpha}V^{e}+T_{~A\beta}^{\alpha}\psi^{\beta}\Big)\iota_{\alpha
}\delta^{4}(\psi)\nonumber\\
&  =\epsilon_{abcd}\Big(T_{~Ae}^{a}V^{e}\Big)\dots V^{d}\delta^{4}%
(\psi)+\epsilon_{abcd}V^{a}\dots V^{d}\Big(T_{~A\beta}^{\alpha}\psi^{\beta
}\Big)\iota_{\alpha}\delta^{4}(\psi)\nonumber
\end{align}
where we have used  $\psi^{\alpha}\delta^{4}(\psi)=0$ and
$V^{1}\wedge\dots\wedge V^{5}=0$. In addition, using $\psi^{\alpha}%
\iota_{\beta}\delta^{4}(\psi)=-\delta_{\beta}^{\alpha}\delta^{4}(\psi)$ and
$V^{a}\wedge\dots\wedge V^{d}=\epsilon^{abcd}(V)^{4}$, we finally find
\begin{equation}
\nabla_{A}\omega^{(4|4)}=(-1)^{B}T_{~~BA}^{B}\omega^{(4|4)}%
\end{equation}
This guarantees, for $T_{~~BA}^{B}=0$ the integration by parts formula
\begin{equation}
\int_{{\cal M}^{(4|4)}} \omega^{(4|4)}\,\nabla_{A}\Phi^{(0|0)}=-\int_{{\cal M}^{(4|4)}}   (\nabla_{A}\omega^{(4|4)}%
)\Phi^{(0|0)} =0 \label{teo1}%
\end{equation}
for a superfield $\Phi^{(0|0)}$.

Now we consider again the top integral form $\omega^{(4|4)}$ and we express it
in terms of curved coordinates as
\begin{align}
\omega^{(4|4)} &  =\Big(\epsilon_{abcd}V^{a}\dots V^{b}\delta^{4}%
(\psi)\Big)\label{theoD}\\
&  =(\epsilon_{abcd}E_{m}^{a}\dots E_{p}^{b})(\epsilon_{\alpha\beta
\gamma\delta}E_{\mu}^{\alpha}\dots E_{\sigma}^{\delta})dx^{m}\dots
dx^{p}\,\delta(d\theta^{\mu})\dots\delta(\delta\theta^{\sigma})\nonumber\\
&  =E\,d^{4}x\delta^{4}(d\theta)\,.\nonumber
\end{align}
where $E=\mathrm{Sdet}(E_{M}^{A})$ is the superdeterminant of the
supervielbein. $E$ is a function of $(x,\theta,\bar{\theta})$ using the
chiral/anti-chiral decomposition. Then, we can expand it according to $\theta$
or $\bar{\theta}$ as follows%
\begin{equation}
\omega^{(4|4)}=E\,d^{4}x\delta^{4}(d\theta)=\Big(\left.  E\right\vert
_{\bar{\theta}=0}+\bar{\theta}^{\dot{\alpha}}\left.  \bar{D}_{\dot{\alpha}%
}E\right\vert _{\bar{\theta}=0}+\bar{\theta}^{2}\left.  \bar{D}^{2}%
E\right\vert _{\bar{\theta}=0}\Big)d^{4}x\delta^{4}(d\theta)+\mathrm{h.c.}%
\label{theoE}%
\end{equation}

Using (\ref{theoD}), we can set
\begin{equation}
\omega^{(4|4)}=E\,d^{4}x\delta^{4}(d\theta)=d\Xi+\left.  \bar{D}%
^{2}E\right\vert _{\bar{\theta}=0}d^{4}x\delta^{2}(d\theta)\bar{\theta}%
^{2}\delta^{2}(d\bar{\theta})+\mathrm{h.c.}\label{theoF}%
\end{equation}
where the first and the second
terms in the expansion in eq. (\ref{theoE}) are cohomologically trivial, while
the third term provides the factor $\bar{\theta}^{2}$ needed to construct the
PCO. In eq. (\ref{theoF}) we have collected the exact terms into $d\Xi$. 

Looking at the superdeterminant $\mathrm{Sdet}(E)$, by choosing a gauge
such that $E_{\dot{\mu}}^{a}=0$ (no mixing between the chiral and the
anti-chiral representation), we have:
\begin{equation}
\mathrm{Sdet}(E)=\frac{\mathrm{det}\Big(E_{m}^{a}-E_{\mu}^{a}(E^{-1})_{\beta
}^{\mu}E_{m}^{\beta}-E_{\dot{\mu}}^{a}(\bar{E}^{-1})_{\dot{\beta}}^{\dot{\mu}%
}E_{m}^{\dot{\beta}}\Big)}{\mathrm{det}(E_{\mu}^{a})\mathrm{det}(\bar{E}%
_{\dot{\mu}}^{\dot{\alpha}})}=\frac{\mathrm{Sdet}_{C}(\hat{E})}{\mathrm{det}%
(\bar{E}_{\dot{\mu}}^{\dot{\alpha}})}\label{theoG}%
\end{equation}
where $\mathrm{Sdet}_{C}(\hat{E})$ is the chiral super determinant written in
terms of a redefined vielbein $\hat{E}_{m}^{a}=E_{m}^{a}-E_{\dot{\mu}}%
^{a}(\bar{E}^{-1})_{\dot{\beta}}^{\dot{\mu}}E_{m}^{\dot{\beta}}$. It can be
proved that, by a suitable gauge fixing (chiral representation) $\mathrm{Sdet}%
_{C}(\hat{E})$ is chiral, namely $\bar{D}_{\dot{\alpha}}\mathrm{Sdet}_{C}%
(\hat{E})=0.$ We can than rewrite the above expression as follows:
\begin{align}
\omega^{(4|4)} &  =E\,d^{4}x\delta^{4}(d\theta)=d\Omega+\mathrm{Sdet}_{C}%
(\hat{E})\left.  \bar{D}^{2}\left(  \frac{1}{\mathrm{det}(\bar{E}_{\dot{\mu}%
}^{\dot{\alpha}})}\right)  \right\vert _{\bar{\theta}=0}d^{4}x\delta
^{2}(d\theta)\bar{\theta}^{2}\delta^{2}(d\bar{\theta})+\mathrm{h.c.}%
\label{theoH}\\
&  =\omega^{(4|2)}\left.  \bar{D}^{2}\left(  \frac{1}{\mathrm{det}(\bar
{E}_{\dot{\mu}}^{\dot{\alpha}})}\right)  \right\vert _{\bar{\theta}=0}%
\bar{\theta}^{2}\delta^{2}(d\bar{\theta})\nonumber
\end{align}
and using the notations of \cite{GGRS} we set $\mathcal{R}=\left.  \bar{D}%
^{2}\left(  \mathrm{det}(\bar{E}_{\dot{\mu}}^{\dot{\alpha}})\right)
^{-1}\right\vert _{\bar{\theta}=0}$. The superfield $\mathcal{R}$ contains the
auxiliary fields and the Ricci scalar and it appears in the commutation
relation $\{\nabla_{\alpha},\nabla_{\beta}\}=-\bar{\mathcal{R}}\mathcal{M}%
_{\alpha\beta}$, namely it is one of the components of the torsion $T^{A}$.

Finally, recalling that $\omega^{(4|2)}=\mathrm{Sdet}_{C}(\hat{E})d^{4}%
x\delta^{2}(d\theta)$ we have:
\begin{equation}
\omega^{(4|4)}=\frac{1}{\mathcal{R}}\omega^{(4|2)}\wedge{\mathbb{Y}}%
^{(0|\bar{2})}+\mathrm{h.c.}\label{teo2}%
\end{equation}
which reproduces Siegel chiral integration formula in terms of
integral forms \cite{GGRS}.

\section*{Acknowledgement} 
We thank P. Fr\'e and C. Maccaferri for fruitful discussions.

\vfill
\eject

\section{Appendix A: gamma matrix conventions and two-component formalism}

\noindent\textbf{Clifford algebra}
\begin{equation}
\{ \gamma_{a}, \gamma_{b} \} = 2 \eta_{ab},~~~~~~~~~ \eta_{ab} = (1,-1,-1,-1)
\end{equation}

\noindent\textbf{Matrix representation}
\begin{equation}
\gamma_{0}= \left(
\begin{array}
[c]{cc}%
0 & 1_{2 \times2}\\
1_{2 \times2} & 0\\
&
\end{array}
\right)  , ~~~\gamma_{i=1,2,3} = \left(
\begin{array}
[c]{cc}%
0 & \sigma_{i}\\
-\sigma_{i} & 0\\
&
\end{array}
\right)  ,~~~\gamma_{5} = -i \gamma_{0} \gamma_{1} \gamma_{2} \gamma_{3} =
\left(
\begin{array}
[c]{cc}%
1_{2 \times2} & 0\\
0 & - 1_{2 \times2}\\
&
\end{array}
\right)
\end{equation}
where $\sigma_{i}$ are the Pauli matrices. The Weyl projectors $P_{\pm}= (1
\pm\gamma_{5} )/2$ are therefore given by
\begin{equation}
P_{+} = \left(
\begin{array}
[c]{cc}%
1_{2 \times2} & 0\\
0 & 0\\
&
\end{array}
\right)  , ~~~~P_{-} = \left(
\begin{array}
[c]{cc}%
0 & 0\\
0 & 1_{2 \times2}\\
&
\end{array}
\right)
\end{equation}
\vskip 2mm

\noindent\textbf{Two-component formalism}

\noindent The four dimensional spinor index is decomposed into $\alpha$=1,2,
${\dot\alpha}$=1,2. Thus
\begin{equation}
\gamma_{0}= \left(
\begin{array}
[c]{cc}%
0 & \delta^{\alpha}_{{\dot\beta}}\\
\delta^{{\dot\alpha}}_{\beta} & 0\\
&
\end{array}
\right)  , ~~~\gamma_{i=1,2,3} = \left(
\begin{array}
[c]{cc}%
0 & {\sigma}_{i~~{\dot\beta}}^{~\alpha}\\
-\sigma_{i~~\beta}^{~{\dot\alpha}} & 0\\
&
\end{array}
\right)  ,~~~\gamma_{5} = \left(
\begin{array}
[c]{cc}%
\delta^{\alpha}_{\beta} & 0\\
0 & - \delta^{{\dot\alpha}}_{{\dot\beta}}\\
&
\end{array}
\right)
\end{equation}
A four-component spinor gets decomposed into two two-component spinors $\psi=
(\psi_{+}^{\alpha}, \psi_{-} ^{{\dot\alpha}})$, where the $\pm$ subscripts
remind us that they are the $P_{\pm}$ projected parts of $\psi$. These
subscripts may be omitted when the $\alpha$ or ${\dot\alpha}$ indices suffice
to identify $\psi_{+}$ or $\psi_{-}$.

\noindent A compact way to express $\gamma_{a=0,1,2,3}$ is
\begin{equation}
\gamma_{a}= \left(
\begin{array}
[c]{cc}%
0 & {\sigma}_{a~~{\dot\beta}}^{~\alpha}\\
-\sigma_{a~~\beta}^{~{\dot\alpha}} & 0\\
&
\end{array}
\right)  ,~~~\mathrm{with} ~~ {\sigma}_{a~~{\dot\beta}}^{~\alpha} = ( 1,
\sigma_{i})_{~~{\dot\beta}}^{~\alpha},~~~ {\sigma}_{a~~\beta}^{~{\dot\alpha}}
= ( -1, \sigma_{i})_{~~\beta}^{~{\dot\alpha}}%
\end{equation}
The matrices $\sigma_{a}$ satisfy the completeness and the trace relations
\begin{equation}
\eta^{ab} {\sigma}_{a~~{\dot\beta}}^{~\alpha} ~{\sigma}_{b~~\delta}%
^{~{\dot\gamma}}= 2~ \delta^{\alpha}_{\delta}~\delta^{{\dot\gamma}}%
_{{\dot\beta}},~~~~ Tr(\sigma_{a} \sigma_{b})= 2 \eta_{ab}%
\end{equation}

\vskip 2mm

\noindent\textbf{Charge conjugation}

\noindent In the above matrix representation, the charge conjugation takes the
form
\begin{equation}
C= \left(
\begin{array}
[c]{cc}%
\epsilon_{\alpha\beta} & 0\\
0 & - \epsilon_{{\dot\alpha}{\dot\beta}}\\
&
\end{array}
\right)
\end{equation}
where $\epsilon$ is the usual Levi-Civita symbol in two dimensions. One can
check that
\begin{equation}
\gamma_{a}^{T} = - C \gamma_{a} C^{-1}%
\end{equation}
so that $C \gamma_{a}$, $C \gamma_{ab}$ are symmetric, while $C$, $C
\gamma_{5}$, $C \gamma_{ab} \gamma_{5}$ are antisymmetric. \vskip 2mm
\noindent\textbf{Majorana condition}

\noindent We can impose the Majorana condition on the spinor $\psi$:
\begin{equation}
\psi^{\dagger}\gamma_{0} = \psi^{T} C
\end{equation}
relating $\psi_{+}^{\alpha}, \psi_{-} ^{{\dot\alpha}}$ to the components of
the conjugated spinor $(\psi^{*}_{+})_{\alpha}, (\psi^{*}_{-})_{{\dot\alpha}}$
as follows:
\begin{equation}
\psi_{+}^{\alpha}\epsilon_{\alpha\beta} = (\psi^{*}_{-})_{\beta},~~~\psi
_{-}^{{\dot\alpha}}\epsilon_{{\dot\alpha}{\dot\beta}} = - (\psi^{*}%
_{+})_{{\dot\beta}}%
\end{equation}
Note that a spinor cannot be both Majorana and Weyl in 4 dimensions, since the
Majorana condition mixes the $\psi_{+}$ and $\psi_{-}$ components. \vskip 2mm
\noindent\textbf{Raising and lowering spinor indices}

\noindent The charge conjugation matrix $C$ and its inverse $C^{-1}$ can be
used to lower and raise spinor indices. Correspondingly $\epsilon_{\alpha
\beta}$ and $\epsilon_{{\dot\alpha}{\dot\beta}}$, and their inverses, are used
to lower and raise two-component spinor indices with the ``upper left to lower
right" convention. Thus for example
\begin{equation}
A_{\alpha}= A^{\beta}\epsilon_{\beta\alpha}, ~~~A^{\alpha}= \epsilon
^{\alpha\beta} A_{\beta}%
\end{equation}
Note that $A^{\alpha}B_{\alpha}= - A_{\alpha}B^{\alpha}$ and similar for
dotted indices. We can also define $\sigma_{a}$ matrices with both indices up
or down:
\begin{equation}
\sigma_{a}^{~\alpha{\dot\beta}} \equiv\epsilon^{{\dot\beta}{\dot\gamma}}
\sigma_{a~~{\dot\gamma}}^{~\alpha},~~~\sigma_{a~\alpha{\dot\beta}}
\equiv\sigma_{a~~{\dot\beta}}^{~\gamma}~\epsilon_{\gamma\alpha},~~~ \sigma
_{a}^{~{\dot\alpha}\beta} \equiv\epsilon^{\beta\gamma} \sigma_{a~~\gamma
}^{~{\dot\alpha}},~~~\sigma_{a~{\dot\alpha}\beta} \equiv\sigma_{a~~\beta
}^{~{\dot\gamma}}~\epsilon_{{\dot\gamma}{\dot\alpha}}%
\end{equation}
With these definitions one finds
\begin{equation}
\sigma_{a}^{~\alpha{\dot\beta}}= \sigma_{a}^{~{\dot\beta}\alpha}%
,~~~\sigma_{a~\alpha{\dot\beta}}=\sigma_{a~{\dot\beta}\alpha}%
\end{equation}
i.e. the $\sigma_{a}$ matrices with both indices up or down are symmetric.
\vskip 2mm \noindent\textbf{Converting vector into spinor indices}

\noindent Finally, the $\sigma_{a}$ matrices can be used to convert a 4-dim
vector index into a couple of two-component spinor indices, and viceversa:
\begin{equation}
V^{\alpha{\dot\alpha}} \equiv V^{a} \sigma_{a}^{~\alpha{\dot\alpha}} ~
\Longrightarrow~ V^{a} = {\frac{1 }{2}} \sigma^{a}_{~\alpha{\dot\alpha}}
V^{\alpha{\dot\alpha}}%
\end{equation}
The second formula can be deduced from the first, and from the trace relation
\begin{equation}
\sigma_{a~\alpha{\dot\alpha}} \sigma_{b}^{~\alpha{\dot\alpha}} = 2 \eta_{ab}%
\end{equation}

\vskip 2mm \noindent\textbf{Examples}

\noindent i) the current $\bar\psi\gamma^{a} \psi$ ($\psi$ Majorana spinor
1-form) becomes, in two-component formalism:
\begin{align}
&  \bar\psi\gamma_{a} \psi= \psi^{T} C \gamma_{a} \psi= \psi^{\alpha}%
\epsilon_{\alpha\beta} \sigma_{a~~{\dot\gamma}}^{~\beta} \psi^{{\dot\gamma}}+
\psi^{{\dot\alpha}}\epsilon_{{\dot\alpha}{\dot\beta}} \sigma_{a~~\gamma
}^{~{\dot\beta}} \psi^{\gamma}=-\psi^{\alpha}\sigma_{a~\alpha{\dot\gamma}}
\psi^{{\dot\gamma}}- \psi^{{\dot\alpha}}\sigma_{a~{\dot\alpha}\gamma}
\psi^{\gamma}=\\
&  = -\psi^{\alpha}\sigma_{a~\alpha{\dot\gamma}} \psi^{{\dot\gamma}}-
\psi^{{\dot\gamma}}\sigma_{a~{\dot\gamma}\alpha} \psi^{\alpha}= - 2
\psi^{\alpha}\psi^{{\dot\gamma}}\sigma_{a~\alpha{\dot\gamma}}%
\end{align}
having used $\sigma_{a~{\dot\gamma}\alpha}=\sigma_{a~\alpha{\dot\gamma}}$.
Converrting the vector index into two-component spinor indices yields:
\begin{equation}
\bar\psi\gamma^{a} \psi~ \sigma_{a}^{~\beta{\dot\delta}} = -2 \psi^{\alpha
}\psi^{{\dot\gamma}}\sigma^{a}_{~\alpha{\dot\gamma}} \sigma_{a}^{~\beta
{\dot\delta}} = 4 \psi^{\beta}\psi^{{\dot\delta}}%
\end{equation}
using the completeness relation. Thus the flat superspace Cartan-Maurer
equation $dV^{a}={\frac{i }{2}} \bar\psi\gamma^{a} \psi$ becomes $d
V^{\alpha{\dot\alpha}} = 2 i \psi^{\alpha}\psi^{{\dot\alpha}}$. \vskip 2mm
\noindent ii) chiral and antichiral projections of $VV$:
\begin{equation}
(V^{2}_{+})^{\alpha\beta} \equiv[P_{+} (VV)]^{\alpha\beta}=V^{a} V^{b} [P_{+}
\gamma_{ab}]^{\alpha\beta}=V^{a} V^{b} \sigma_{a~~{\dot\beta}}^{\alpha}~
\sigma_{b}^{~{\dot\beta}\beta}=V^{a} V^{b} \sigma_{a}^{~\alpha{\dot\alpha}}
\sigma_{b}^{~{\dot\beta}\beta} \epsilon_{{\dot\alpha}{\dot\beta}}%
=V^{\alpha{\dot\alpha}} V^{\beta{\dot\beta}} \epsilon_{{\dot\alpha}{\dot\beta
}}%
\end{equation}
and similarly $(V^{2}_{-})^{{\dot\alpha}{\dot\beta}} =V^{\alpha{\dot\alpha}}
V^{\beta{\dot\beta}} \epsilon_{\alpha\beta}$. \vskip 2mm \noindent\textbf{Pros
and cons}

\noindent Pros: two-component formalism can simplify calculations, since gamma
matrices disappear in most cases (see the examples above), and Fierz
rearrangements are automatically implemented. Cons: the notation is less
compact (two spinor indices replace one vector index), and it is necessary to
remember minus signs in relations like $A^{\alpha}B_{\alpha}= - A_{\alpha
}B^{\alpha}$.


\section{Appendix B: Some useful formulas}

We consider the supervielbeins $(V^{\alpha{\dot{\alpha}}},\psi^{\alpha}%
,\bar{\psi}^{{\dot{\alpha}}})$ such that
\begin{equation}
dV^{\alpha{\dot{\alpha}}}=2i\psi^{\alpha}\bar{\psi}^{{\dot{\alpha}}%
}\,,~~~~~~d\psi^{\alpha}=0\,,~~~~~d\bar{\psi}^{{\dot{\alpha}}}=0\,.
\label{ALA}%
\end{equation}

We define the following combinations
\begin{align}
&  (V_{+}^{2})^{\alpha\beta}=\frac{1}{2!}V^{\alpha{\dot{\alpha}}}\wedge
V^{\beta{\dot{\beta}}}\epsilon_{{\dot{\alpha}}{\dot{\beta}}}%
\,,~~~~~~\label{ALB}\\
&  (V_{-}^{2})^{{\dot{\alpha}}{\dot{\beta}}}=\frac{1}{2!}V^{\alpha{\dot
{\alpha}}}\wedge V^{\beta{\dot{\beta}}}\epsilon_{\alpha\beta}%
\,,~~~~~~\nonumber\\
&  (V^{3})^{\alpha{\dot{\alpha}}}=\frac{1}{3!}V^{\alpha{\dot{\beta}}}\wedge
V^{\dot{\gamma}\gamma}\wedge V^{\beta\dot{\alpha}}\epsilon_{{\dot{\beta}}%
\dot{\gamma}}\epsilon_{\gamma\beta}\,,\nonumber\\
&  (V^{4})=\frac{1}{4!}V^{\alpha{\dot{\alpha}}}\wedge V^{\dot{\gamma}\gamma
}\wedge V^{\beta\dot{\beta}}\wedge V^{\dot{\sigma}\sigma}\epsilon
_{{\dot{\alpha}}\dot{\gamma}}\epsilon_{\gamma\beta}\epsilon_{\dot{\beta}%
\dot{\sigma}}\epsilon_{\sigma\alpha} = \det(V^{\a\dot\a})\,,\nonumber
\end{align}
The first two combinations $V_{\pm}^{2}$ are the self-dual and anti-self dual
part of the wedge product of two vielbeins $V^{\alpha\dot{\alpha}}$. The last
one is the singlet combination (corresponding to the determinant) of the
vielbein. By multiplying with $V^{\alpha{\dot{\alpha}}}$ we find the following
relations (recall that $e_{\alpha\beta}\epsilon^{\beta\gamma}=\delta_{\alpha
}^{\gamma}$ and $e_{\alpha\beta}\epsilon^{\alpha\beta}=-2$)
\begin{align}
&  V^{\alpha\dot{\alpha}}\wedge V^{\beta\dot{\beta}}=-\epsilon^{\alpha\beta
}(V_{-}^{2})^{{\dot{\alpha}}{\dot{\beta}}}-\epsilon^{{\dot{\alpha}}{\dot
{\beta}}}(V_{+}^{2})^{\alpha\beta}\,,\label{ALC} \nonumber \\
&  V^{\alpha{\dot{\alpha}}}\wedge V^{\beta{\dot{\beta}}}\wedge V^{\gamma
{\dot{\gamma}}}=2\epsilon^{{\dot{\beta}}{\dot{\gamma}}}\epsilon^{\alpha(\beta
}(V^{3})^{\gamma){\dot{\alpha}}}+2\epsilon^{\beta\gamma}\epsilon^{{\dot
{\alpha}}({\dot{\beta}}}(V^{3})^{{\dot{\gamma}})\alpha}\,,\nonumber\\
&  V^{\alpha\dot{\alpha}}\wedge(V_{+}^{2})^{\beta\gamma}=-2\epsilon
^{\alpha(\beta}(V^{3})^{\gamma)\dot{\alpha}}\,,\nonumber\\
&  V^{\alpha\dot{\alpha}}\wedge(V_{-}^{2})^{{\dot{\beta}}{\dot{\gamma}}%
}=-2\epsilon^{{\dot{\alpha}}({\dot{\beta}}}(V^{3})^{{\dot{\gamma}})\alpha
}\,,\nonumber\\
&  V^{\alpha\dot{\alpha}}\wedge(V^{3})^{\beta{\dot{\beta}}}=\epsilon
^{\alpha\beta}\epsilon^{{\dot{\alpha}}{\dot{\beta}}}(V^{4})\,,\nonumber\\
&  V^{\alpha{\dot{\alpha}}}\wedge V^{\beta{\dot{\beta}}}\wedge V^{\gamma
{\dot{\gamma}}}\wedge V^{\sigma\dot{\sigma}}=t^{\alpha{\dot{\alpha}}\beta
{\dot{\beta}}\gamma{\dot{\gamma}}\sigma\dot{\sigma}}(V^{4})\,, \nonumber \\
& (V_+^2)^{\a\b} \wedge (V_+^2)^{\g\delta} = 
(\epsilon^{\a\g} \epsilon^{\b \delta} +  \epsilon^{\a\delta} \epsilon^{\b \gamma}) (V^4)\,,\nonumber \\
& (V_-^2)^{\dot\a\dot\b} \wedge (V_-^2)^{\dot\g\dot\delta} = 
(\epsilon^{\dot\a\dot\g} \epsilon^{\dot\b \dot\delta} +  \epsilon^{\dot\a\dot\delta} 
\epsilon^{\dot\b \dot\gamma}) (V^4)\,, \nonumber \\
& (V_+^2)^{\a\b} \wedge (V_-^2)^{\dot\g\dot\delta} =0\,, 
\end{align}
where $A^{(\alpha\beta)}=\frac{1}{2}(A^{\alpha\beta}+A^{\beta\alpha})$ and the
tensor
\begin{align}
t^{\alpha{\dot{\alpha}}\beta{\dot{\beta}}\gamma{\dot{\gamma}}\sigma\dot
{\sigma}} &=
-\epsilon^{\alpha\beta}\epsilon^{{\dot{\beta}}{\dot{\gamma}}}\epsilon^{\gamma\sigma}\epsilon^{\dot{\sigma}{\dot{\alpha}}}
+\epsilon^{\alpha\gamma}\epsilon^{{\dot{\gamma}}{\dot{\beta}}}\epsilon^{\beta\sigma}\epsilon^{\dot{\sigma}{\dot{\alpha}}}
-\epsilon^{\alpha\sigma}\epsilon^{\dot\sigma{\dot{\gamma}}}\epsilon^{\gamma\beta}\epsilon^{{\dot{\beta}}{\dot{\alpha}}}
+\epsilon^{\alpha\sigma}\epsilon^{{\dot{\sigma}}{\dot{\beta}}}\epsilon^{\beta\gamma}\epsilon^{{\dot{\gamma}}{\dot{\alpha}}}\nonumber\\
&
= \epsilon^{\dot\a\dot\b}\epsilon^{\dot\gamma\dot\sigma} \epsilon^{\a\gamma}\epsilon^{\b\sigma} 
+ \epsilon^{\dot\a\dot\b}\epsilon^{\dot\gamma\dot\sigma} \epsilon^{\a\sigma}\epsilon^{\b\gamma} 
+ \epsilon^{\dot\a\dot\gamma}\epsilon^{\dot\beta\dot\sigma} \epsilon^{\a\b}\epsilon^{\gamma\sigma} 
+ \epsilon^{\dot\a\dot\sigma}\epsilon^{\dot\beta\dot\gamma} \epsilon^{\a\b}\epsilon^{\gamma\sigma} 
\end{align}
respects all properties of form multiplication. The second line is obtained 
by using relations like
$$
\epsilon^{\a\b} \epsilon^{\gamma\delta} +
\epsilon^{\a\gamma} \epsilon^{\delta\beta} + 
\epsilon^{\a\delta} \epsilon^{\beta\gamma} =0\,.
$$
This invariant tensor is
obtained by contracting with the Dirac gamma matrices the Levi-Civita tensor
\begin{equation}
t^{\alpha{\dot{\alpha}}\beta{\dot{\beta}}\gamma{\dot{\gamma}}\sigma\dot
{\sigma}}=\frac{1}{4!}\epsilon_{abcd}(\gamma^{a})^{\alpha{\dot{\alpha}}%
}(\gamma^{b})^{\beta{\dot{\beta}}}(\gamma^{c})^{\gamma{\dot{\gamma}}}%
(\gamma^{d})^{\sigma\dot{\sigma}}\,. \label{ALDA}%
\end{equation}

The differentials are
\begin{align}
&  dV^{\alpha{\dot{\alpha}}}=2i\psi^{\alpha}\bar{\psi}^{{\dot{\alpha}}%
}\,,~~~~~\label{ALE}\\
&  d(V_{+}^{2})^{\alpha\beta}=-2i\psi^{(\alpha}(V\bar{\psi})^{\beta
)}\,,\nonumber\\
&  d(V_{-}^{2})^{{\dot{\alpha}}{\dot{\beta}}}=-2i\bar{\psi}^{({\dot{\alpha}}%
}(V\psi)^{{\dot{\beta}})}\,,\nonumber\\
&  d(V^{3})^{\alpha{\dot{\alpha}}}=i\Big(\psi^{\alpha}(V_{-}^{2}\bar{\psi
})^{{\dot{\alpha}}}-\bar{\psi}^{{\dot{\alpha}}}(V_{+}^{2}\psi)^{\alpha
}\Big)\,,\nonumber\\
&  d(V^{4})=2i\psi_{\alpha}(V^{3})^{\alpha\dot{\alpha}}\bar{\psi}%
_{{\dot{\alpha}}}\,,\nonumber
\end{align}
where
\begin{align}
&  (V\bar{\psi})^{\alpha}=V^{\alpha{\dot{\alpha}}}\epsilon_{{\dot{\alpha}%
}{\dot{\beta}}}\,\bar{\psi}^{{\dot{\beta}}}\,,\label{ALF}\\
&  (V\psi)^{{\dot{\alpha}}}=V^{{\dot{\alpha}}\alpha}\epsilon_{\alpha\beta
}\,\psi^{\beta}\,,\nonumber\\
&  (V_{-}^{2}\bar{\psi})^{{\dot{\alpha}}}=(V_{-}^{2})^{{\dot{\alpha}}%
{\dot{\beta}}}\epsilon_{{\dot{\beta}}{\dot{\gamma}}}\,\bar{\psi}^{{\dot
{\gamma}}}\,,\nonumber\\
&  (V_{+}^{2}\psi)^{\alpha}=(V_{+}^{2})^{\alpha\beta}\epsilon_{\beta\gamma
}\,\psi^{\gamma}\,,\nonumber
\end{align}
with 
\begin{align}
& d (V\bar{\psi})^{\alpha}= 2 i \psi^\a \bar\psi^{\dot \a}
\epsilon_{{\dot{\alpha}}{\dot{\beta}}}\,\bar{\psi}^{{\dot{\beta}}} =0 \,,\label{ALF} \nonumber \\
&  d (V\psi)^{{\dot{\alpha}}}=
2 i \psi^\a \bar\psi^{\dot \a} \epsilon_{\alpha\beta}\,\psi^{\beta} =0 \,,\nonumber\\
&  d (V_{-}^{2}\bar{\psi})^{{\dot{\alpha}}}= i \bar \psi^{\dot\a} (\psi V \bar \psi)
\,,\nonumber\\
&  d (V_{+}^{2}\psi)^{\alpha}=  i \psi^\a (\psi V \bar\psi)\,.\nonumber \\
& d (\bar \psi V_{-}^{2}\bar{\psi}) = 0\,, \nonumber \\
& d (\psi V_{+}^{2} {\psi}) = 0\,, \nonumber \\
& d (\psi V \bar \psi) =0\,. 
\end{align}

\section{Appendix C: the curved supermanifold $Osp(1|4)$.}

Let us consider the case of curved supermanifolds, for example the supercoset manifold
\[
\mathrm{Osp}(1|4)/\mathrm{SO(1,3)}\sim(\mathrm{AdS}_{4}|4)
\]
which is a supermanifold whose bosonic submanifold is $4d$ anti-de Sitter and
with $4$ fermionic coordinates. We have
\begin{align}
&  \nabla V^{\alpha\dot{\alpha}}=2 i \psi^{\alpha}\wedge\bar{\psi}%
^{\dot{\alpha}}\,,~~~\\
&  \nabla\psi^{\alpha}=i\Lambda\,V^{\alpha\dot{\alpha}}\bar{\psi}_{\dot
{\alpha}}\,,~~~\\
&  \nabla\bar{\psi}^{\dot{\alpha}}=-i \Lambda\,V^{\alpha\dot{\alpha}}%
\psi_{\alpha}\,.\label{SupE}\\
&  R^{\alpha\beta}_{+} = 4 \Lambda^{2} (V^{2}_{+})^{\alpha\beta} + 2
\Lambda\psi^{\alpha}\psi^{\beta}\,, ~~~~~\\
&  R^{{\dot\alpha}{\dot\beta}}_{-} = - 4 \Lambda^{2} (V^{2}_{-})^{{\dot\alpha
}{\dot\beta}} - 2 \Lambda\bar\psi^{{\dot\alpha}}\bar\psi^{{\dot\beta}}\,,
~~~~~
\end{align}
where $R^{ab}$, decomposed into the self-dual and the anti-self dual parts
($R^{\alpha\beta}_{+}$ and $R^{{\dot\alpha}{\dot\beta}}_{-}$), is the curvature of
the supermanifold and $\nabla$ is the covariant derivative w.r.t. to the
connection of $\mathrm{SO}(1,3)$. It iseasy to check the Bianchi identities
using these definitions.

In addition, by using (\ref{SupE}), we can verify that
\begin{align}
d\left(  (V^{4})\delta^{4}(\psi)\right)   &  =\nabla\left(  (V^{4})\delta
^{4}(\psi)\right) \label{SupF}\\
&  =2i\psi_{\alpha}(V^{3})^{\alpha{\dot{\alpha}}}\bar{\psi}_{{\dot{\alpha}}%
}\delta^{4}(\psi)+(V^{4})i\Lambda V^{\alpha{\dot{\alpha}}}\left(  \bar{\psi
}_{{\dot{\alpha}}}\iota_{\alpha}-\psi_{\alpha}\bar{\iota}_{{\dot{\alpha}}%
}\right)  \delta^{4}(\psi)\,.\nonumber
\end{align}
where the first equality follows from the Lorentz invariance of the volume
form $V^{4}\delta^{4}(\psi)$ and the last follows from the distributional law
$\psi\delta(\psi)=0$ and the properties of top forms.

For the curved case, we have that
\begin{equation}
\nabla\Big(V^{a}\bar{\psi}\gamma_{a}\psi\Big)=i\Lambda V^{a}\wedge
V^{b}\big(\psi\gamma_{ab}\psi-\bar{\psi}\gamma_{ab}\bar{\psi}\Big)
\end{equation}
and since $\Big(V^{a}\bar{\psi}\gamma_{a}\psi\Big)$ is a scalar, we find 
$\nabla^{2}\Big(V^{a}\bar{\psi}\gamma_{a}\psi\Big)=0$. This means that only
the class $iV^{a}\wedge V^{b}\big(\psi\gamma_{ab}\psi-\bar{\psi}\gamma
_{ab}\bar{\psi}\Big)$ is closed. In the limit $\Lambda\rightarrow0$, one recovers
the flat case.

Let us now consider the same problem in the curved space. We start with
$\mathrm{Osp}(1|4)$ case. We use the relations given in (\ref{SupE}) and the
volume form has the expression
\begin{equation}
\omega^{(4|4)}=\epsilon_{abcd}V^{a}\wedge\dots\wedge V^{d}\wedge\delta
^{4}(\psi)\,,
\end{equation}
which is closed (the variation of $V^{a}$ is cancelled because of the
Dirac delta's, while the variation of $\psi$'s is cancelled by the presence of
four $V$'s. Using the definitions
\begin{equation}
V^{a}=V_{m}^{a}dx^{m}+V_{\mu}^{a}d\theta^{\mu}\,,~~~~~~\psi^{a}=\psi_{m}%
^{a}dx^{m}+\psi_{\mu}^{a}\theta^{\mu}\,,
\end{equation}
we find
\begin{equation}
\omega^{(4|4)}=\mathrm{Sdet}(E)\epsilon_{abcd}dx^{a}\wedge\dots\wedge
dx^{d}\delta^{4}(d\theta)
\end{equation}
with $E=\left(
\begin{array}
[c]{cc}%
V_{m}^{a} & V_{\mu}^{a}\\
\psi_{m}^{\alpha} & \psi_{\mu}^{\alpha}%
\end{array}
\right)  $. The bosonic space is $\mathrm{Sp}(4)/\mathrm{SO(1,3)}$, namely the
curved space $AdS_{4}$, and therefore we have
\begin{equation}
\mathrm{Vol}_{\mathrm{Osp}(1|4)/\mathrm{SO(1,3)}}=\int_{AdS_{4}}d^{4}x\left.
D^{4}\mathrm{Sdet}(E)\right\vert _{\theta=0}%
\end{equation}
where $D^{4}=\epsilon_{\alpha_{1}\dots\alpha_{4}}D^{\alpha_{1}}\dots
D^{\alpha_{4}}$. In the present case the $(4|4)$-integral form $\omega
^{(4|4)}$ is closed, but it is not exact since $\mathrm{Sdet}(E)$ has a
non-trival $\theta$-dependence.



\begin{thebibliography}{99}              

\bibitem {Castellani:2014goa}L.~Castellani, R.~Catenacci and P.~A.~Grassi,
\textit{Supergravity Actions with Integral Forms,} Nucl.\ Phys.\ B
\textbf{889}, 419 (2014) [arXiv:1409.0192 [hep-th]].

\bibitem {Castellani:2015paa}L.~Castellani, R.~Catenacci and P.~A.~Grassi,
\textit{The Geometry of Supermanifolds and new Supersymmetric Actions},
Nucl.\ Phys.\ B \textbf{899}, 112 (2015) [arXiv:1503.07886 [hep-th]].

\bibitem {Castellani:2015ata}L.~Castellani, R.~Catenacci and P.~A.~Grassi,
\textit{Hodge Dualities on Supermanifolds}, Nucl.\ Phys.\ B \textbf{899}, 570
(2015) [arXiv:1507.01421 [hep-th]].

\bibitem {berkovits}
  N.~Berkovits,
\textit{Multiloop amplitudes and vanishing theorems using the pure spinor formalism for the superstring},
  JHEP {\bf 0409}, 047 (2004)
  doi:10.1088/1126-6708/2004/09/047
  [hep-th/0406055].

\bibitem {Witten:2012bg}E.~Witten, \textit{Notes on Supermanifolds and
Integration,} [arXiv:1209.2199 [hep-th]].
                                                                                 
\bibitem{voronov-book} T. Voronov, \textit{Geometric integration theory on
supermanifolds}. Soviet Scientific Review, Section C: Mathematical Physics,
9, Part 1. Harwood Academic Publisher, Chur. 1991 (Second edition 2014).

\bibitem {Castellani:2017ycm}L.~Castellani, R.~Catenacci and P.~A.~Grassi,
\textit{Super Quantum Mechanics in the Integral Form Formalism}, arXiv:1706.04704 [hep-th].

\bibitem{d2} L. Castellani, R. Catenacci, P.A. Grassi, \textit{Sigma models and chiral bosons in integral superspace}, in preparation.

\bibitem{3dsuper} L. Castellani, R. Catenacci, P.A. Grassi, JHEP10 (2016)
049, \textit{The integral form of supergravity, } [arXiv:1607.05193 [hep-th]]

\bibitem{baggerwess} J. Wess, J. Bagger, \textit{Supersymmetry and Supergravity}%
, Princeton Univ. Press; 2 Revised edition (1992).

\bibitem {GGRS}S.~J.~Gates, Jr., M.~T.~Grisaru, M.~Ro\v{c}ek and W.~Siegel,
\textit{Superspace, or One Thousand and One Lessons in Supersymmetry},
Front.\ Phys.\ \textbf{58}, 1 (1983) [arXiv:hep-th/0108200].

\bibitem {Castellani}L.~Castellani, R.~D'Auria and P.~Fr\'e, \textit{Supergravity
and superstrings: A Geometric perspective}, in 1,2,3 vol., Singapore: World Scientific (1991);

\bibitem{Grimm:1977xp} 
  R.~Grimm, M.~Sohnius and J.~Wess,
  \textit{Extended Supersymmetry and Gauge Theories},
  Nucl.\ Phys.\ B {\bf 133}, 275 (1978).
  doi:10.1016/0550-3213(78)90303-6

\bibitem{AlvarezGaume:1996mv} 
  L.~Alvarez-Gaume and S.~F.~Hassan,
  \textit{Introduction to S duality in N=2 supersymmetric gauge theories: A Pedagogical review of the work of Seiberg and Witten},
  Fortsch.\ Phys.\  {\bf 45}, 159 (1997)
  doi:10.1002/prop.2190450302
  [hep-th/9701069].

\bibitem {LMP}L.Castellani, R.Catenacci, P.A.Grassi, \textit{Integral
Representations on Supermanifolds: super Hodge duals, PCOs and Liouville
forms}, Lett. Math. Phys. (2016). doi:10.1007/s11005-016-0895-x,
[arXiv:1603.01092 [hep-th]]


\bibitem {Catenacci:2010cs}R.~Catenacci, M.~Debernardi, P.~A.~Grassi and
D.~Matessi, \textit{Cech and de Rham Cohomology of Integral Forms},''
J.\ Geom.\ Phys.\ \textbf{62} (2012) 890 doi:10.1016/j.geomphys.2011.12.011
[arXiv:1003.2506 [math-ph]].

\bibitem{Buchbinder:1998qv} 
  I.~L.~Buchbinder and S.~M.~Kuzenko,
 \textit{Ideas and methods of supersymmetry and supergravity: Or a walk through superspace},
  Bristol, UK: IOP (1998) 656 p


\end{thebibliography}
\end{document}